\documentclass[prb,preprint,preprintnumbers,amsmath,amssymb,floats]{revtex4}
\usepackage{graphicx}
\usepackage{dcolumn}
\usepackage{bm,epstopdf,soul}

\usepackage{amsmath}
\usepackage{color}
\usepackage{soul}

\begin{document}
\preprint{}


\title{Stable topological insulators achieved using high energy electron beams}


\small
\author{Lukas Zhao,$^{1}$ Marcin Konczykowski,$^2$ Haiming Deng,$^1$ Inna Korzhovska,$^1$ Milan Begliarbekov,$^1$ Zhiyi Chen,$^1$ Evangelos Papalazarou,$^3$ Marino Marsi,$^3$
Luca Perfetti,$^2$ Andrzej Hruban,$^4$ Agnieszka Wo{\l}o\'{s},$^{5,6}$ \& Lia Krusin-Elbaum$^{1}$}
\vspace{10mm}
\affiliation{$^1$Department of Physics, The City College of New York, CUNY, New York, NY 10031, USA}
\affiliation{$^2$Laboratoire des Solides Irradi\'{e}s, \'{E}cole Polytechnique, CNRS, CEA, Universit\'{e} Paris-Saclay, 91128 Palaiseau cedex, France}
\affiliation{$^3$Laboratoire de Physique des Solides, CNRS, Universit\'{e} Paris-Saclay, Universit\'{e} Paris-Sud, 91405 Orsay, France}%
\affiliation{$^4$Institute of Electronic Materials Technology, 01-919 Warsaw, Poland}%
\affiliation{$^5$Institute of Physics, Polish Academy of Sciences, 02-668 Warsaw, Poland}
\affiliation{$^6$Faculty of Physics, University of Warsaw, 00-681 Warsaw, Poland}

\begin{abstract}
\vspace{10mm}

\noindent \textbf{Topological insulators are transformative quantum solids with immune-to-disorder metallic surface states having Dirac band structure. Ubiquitous charged bulk defects, however, pull the Fermi energy into the bulk bands, denying access to surface charge transport. Here we demonstrate that irradiation with swift ($\sim 2.5$ MeV energy) electron beams allows to compensate these defects, bring the Fermi level back into the bulk gap, and reach the charge neutrality point (CNP). Controlling the beam fluence we tune bulk conductivity from \textit{p}- (hole-like) to \textit{n}-type (electron-like), crossing the Dirac point and back, while preserving the Dirac energy dispersion. The CNP conductance has a two-dimensional (2D) character on the order of ten conductance quanta $G_0 =e^2/h$, and reveals, both in Bi$_2$Te$_3$ and Bi$_2$Se$_3$, the presence of only two quantum channels corresponding to two topological surfaces. The intrinsic quantum transport of the topological states is accessible disregarding the bulk size.
 }
\end{abstract}

\maketitle
\normalsize

Unconventional quantum matter can be easily hidden within the rich existing
library of condensed matter and to recognize it often novel concepts have to be invoked.
It also takes a profound understanding of the real material constraints that can prevent the unconventional properties from being detected.
Three-dimensional (3D) topological insulators  are a spectacular example of this \cite{Transformative2010} -- narrow-band semiconductors, well known for their high performance as thermoelectrics, \cite{ThermoelectricTI2001} they were discovered to support unusual gapless robust two-dimensional (2D) surface states that are fully spin-polarized with Dirac-type linear electronic energy-momentum dispersion \cite{Fu2007,Qi2011}, which makes them protected against backscattering by disorder \cite{Fu2007,Qi2011,Zhang-NatPhys09,Hsieh2009}. These materials have narrow bulk gaps ($\sim 200-300$ meV) and have charge carriers donated by intrinsic crystalline lattice defects such as vacancies and antisites \cite{Scanlon-antisites2012}. As a result, the conduction through the bulk and its intermixing with surface channels is what largely denies direct access to surface charge transport
required for the implementation in spin-based nanoelectronics \cite{Moore-PRL2010} or fault-tolerant topological quantum computing \cite{Qi2011}.

Extensive attempts to reduce the contribution of bulk carriers, involving techniques such as nanostructured synthesis/growth \cite{Cui-nanoplates2010,Oh2014}, chemical doping \cite{Ca-dopingOng2009} or compositional tuning \cite{Zhang-ternary2011},
have relied on electrostatic gating of micro- or nano-structures comprising tens of nanometer thin films \cite{Chen-gating2010} or similarly thin exfoliated crystals \cite{Fuhrer-gating2012}
to gain less ambiguous access to the surface states.
Here we demonstrate that bulk conductivity in topological insulators (TIs) can be decreased by orders of magnitude to charge neutrality point on a large (depth) scale by the controlled use of electron beams which for energies below $\sim 3~ \textrm{MeV}$ are known to produce a well-defined, stable and uniform spread of Frenkel (vacancy-interstitial) pairs \cite{BoisBeuneu1988}
within their penetration range of hundreds of microns
The combined effect of these pairs is to compensate for the intrinsic charged defects responsible for the conductivity of the bulk while crystal lattice integrity is maintained and, as we demonstrate by angularly resolved photoemission spectroscopy (ARPES), the topological Dirac surface states remain robust.
Stable surface conduction channels are achieved when a sufficient irradiation dose is followed by the optimally engineered annealing protocol, thereby resolving one of the key limitations of bulk TIs.

\vspace{3mm}
\noindent\textbf{Results}\\
\noindent\textbf{Compensating bulk carriers using electron beams.}
Under exposure to light particles such as electrons the production of vacancy-interstitial point defects \cite{Damask-Dienes1963} in solids known as Frenkel pairs is well established.
The pair formation has an energy threshold which scales with atomic weight -- it is high for heavy Bi and is lower for lighter Te and Se, and the effective cross-section $\sigma$ for pair creation on different sublattices in materials containing these elements depends on the projectile energy ({Figs.~1a,b and Supplementary Figs.~1,2}). This defines a pair-production window: below these thresholds the given sub-lattice becomes immune to irradiation while at energies above $\sim 3$ MeV clustering and even nuclear reactions \cite{Nuclear2007} may occur.
The process of Frenkel pair formation is  straightforward and can be well controlled. The charge is primarily delivered by introduced vacancies, since at room temperature interstitials do not contribute owing to their much lower (compared to vacancies) migration barriers.
For electron beams energies above $\sim 1.5~\textrm{MeV}$ the effective cross-section $\sigma$ on a Bi sublattice is the highest and we will show that donor type defects on this sublattice do prevail. For energies below Bi threshold ($\sim 1.2~\textrm{MeV}$) the creation of Frenkel pairs will be mostly on Se or Te sublattices and, correspondingly, by tuning electron beam energy a different mix of donor/acceptor defects and a net acceptor charge doping can be expected.

Generally, crystal growth of tetradymite crystals such as Bi$_2$Te$_3$ and Bi$_2$Se$_3$ results in equilibrium defect configurations comprising vacancies and antisites on both sublattices \cite{Scanlon-antisites2012,Xue-defects2013}. The net free charge balance that delivers carriers to bulk conduction or valence bands from these defects can be varied by growth conditions or doping \cite{Ca-dopingOng2009}. In undoped Bi$_2$Se$_3$, where Se vacancies are presumed to dominate, the net conduction is by electron carriers or \textit{n}-type.  In Bi$_2$Te$_3$, where antisites are prevalent \cite{Drasar2010} the conductivity is usually \textit{p}-type, namely by hole carriers, although by varying stoichiometry it has been grown of either conductivity type. Without \textit{a priori} knowledge of the net donor or acceptor flavor of the pairs on different sublattices we have chosen first to irradiate \textit{p}-type topological materials Bi$_2$Te$_3$ and  Ca-doped \cite{Ca-dopingOng2009} Bi$_2$Se$_3$ with 2.5 MeV electron beams which create Frenkel pairs on all sublattices according to $\sigma(E)$, {shown in Fig.~1b and Supplementary Fig.~2.} {Our results demonstrate that in the above mentioned Bi-based TIs at this energy the net flavor is of donor type. It is also important to note that even for relatively high ($\gtrsim 1 ~\textrm{C}\cdot \textrm{cm}^{-2}$) electron irradiation doses the lattice parameters remain unchanged (see Supplementary Fig.~3).}

\vspace{3mm}
\noindent\textbf{Ambipolar conduction tuned by irradiation dose.} Longitudinal resistivity $\rho_{\textrm{xx}}$ of the initially \textit{p-}type Bi$_2$Te$_3$ measured at 20 K \textit{in situ} in the irradiation chamber (see Methods)
as a function of irradiation dose is shown in {Fig.~1e}. Most immediately notable features in the figure are (i) a nearly three orders of magnitude resistivity increase to a maximum $\rho_{\textrm{xx}}^{\textrm{max}}$ and (ii) the observed ambipolar conduction as a function of irradiation dose with well-distinguished \textit{p} (hole) and \textit{n}(electron) conduction regions.
The resistivity maximum $\rho_{\textrm{xx}}^{\textrm{max}}$ (the conductivity minimum $\sigma_{\textrm{xx}}^{\textrm{min}}$) is at the charge neutrality point where conduction is converted from \textit{p-} to \textit{n-}type, as determined from Hall resistivity (see Fig. 2). The same type conversion is observed in Ca-doped Bi$_2$Se$_3$ {(Supplementary Fig.~4a and Supplementary Note 1)} which is also \textit{p-}type \cite{Ca-dopingOng2009}. As long as the terminal irradiation dose $\phi$ is relatively low, below $\sim 0.1~ \textrm{C} \cdot \textrm{cm}^{-2}$,
$\rho_{\textrm{xx}}(\phi)$ traces its shape upon temperature cycling to room temperature and back to 20 K, with $\rho_{\textrm{xx}}^{\textrm{max}}$ reproduced by the next irradiation cycle. Here it is apparent that the value of $\phi_{\textrm{max}}$ at the charge neutrality point is not universal; it depends on the starting free carrier concentration $n_\textrm{b}$ but can be straightforwardly scaled using universal slope $\simeq \frac{\partial n_\textrm{\textrm{b}}}{\partial \phi}$ of quasi-linear variation of $n_\textrm{b}$ \textit{vs}. dose (Fig.~2a and Supplementary Figs.~4b and 5).

\vspace{3mm}
\noindent\textbf{Longitudinal and Hall resistivities across CNP.}
\textit{Ex situ} measured resistivities of Bi$_2$Te$_3$ crystals irradiated to different terminal doses and taken to room temperature before the chill-down to 4.2 K  are in full correspondence with the \textit{in situ} results. Figure 2a shows three orders of magnitude increase in $\rho_{\textrm{xx}}$ to the maximum, $\rho_{\textrm{xx}}^{\textrm{max}}$, at $\phi_{\textrm{max}}\approx 90~\textrm{C}\cdot \textrm{cm}^{-2}$ higher than the \textit{in situ} $\phi_{\textrm{\textrm{max}}}$ {(Fig.~ 1e)} likely owing in part to some defect migration (mostly interstitials \cite{Damask-Dienes1963}) above 100 K.
The \textit{p-} to \textit{n-}type conversion is clearly seen in Hall resistance $R_{\textrm{xy}}$ (Fig.~2b and Figs. 2f-2h) flipping its slope $\textrm{d}R_{\textrm{xy}}/\textrm{d}H$ and Hall coefficient $R_\textrm{H} = -\frac{1}{n_\textrm{b} e}$ changing sign in the conversion region. Near charge neutrality point (CNP) the net residual bulk carrier density $n_\textrm{b} = n_\textrm{D} - n_\textrm{A} $ is very low ($n_\textrm{D}$ and $n_\textrm{A} $ are concentrations of donors and acceptors respectively). In this region local charge fluctuations can be very large creating inhomogeneity akin to a network of puddle-like \textit{p-n} junctions that nonlinearly screen random potential on long length scales \cite{Shklovskii2012} and a simple estimate of $n_\textrm{b}$ is no longer appropriate \cite{Fuhrer-gating2012}.
Mobilities in our crystals are significantly higher ($\mu \cong 7,000 - 11,000~ \textrm{cm}^2 \textrm{V}^{-1} \textrm{s}^{-1}$, see Supplementary Table~1) than a commonly observed $\lesssim 1,000~ \textrm{cm}^2 \textrm{V}^{-1} \textrm{s}^{-1}$ range \cite{Fuhrer-gating2012} near the CNP region.

\vspace{3mm}
\noindent\textbf{Shubnikov-de Haas oscillations and Berry phase.}
The change in net carrier density induced by irradiation is reflected in the observed Shubnikov-de Haas (SdH) quantum oscillations (Fig. 2c-2e) of sheet and Hall resistances, $\Delta R_{\textrm{xx}}$ and $\Delta R_{\textrm{xy}}$, from which an estimate of the Fermi surface size can be obtained (Supplementary Table~1). The Fermi vector in pristine crystals $k_\textrm{F} \approx 0.025\AA^{-1}$ is consistently larger than after irradiation and, for example, the dose $\phi \cong 89~\textrm{C}\cdot \textrm{cm}^{-2}$ which tunes the crystal close to CNP results in $k_\textrm{F} \approx 0.014\AA^{-1}$. The corresponding net carrier densities are $n_\textrm{b} = 5.06 \times 10^{17}\textrm{cm}^{-3}$ and $n_\textrm{b} \simeq 1.08 \times 10^{17}\textrm{cm}^{-3}$ respectively. At higher irradiation dose $\phi \cong  99~\textrm{C}\cdot \textrm{cm}^{-2}$, with $k_\textrm{F} \approx 0.022\AA^{-1}$ Fermi surface size becomes comparable to the pristine material.
A comparison of these numbers with carrier densities obtained independently from Hall data (Supplementary Table~1) reveals a good agreement between the two techniques near CNP but about an order of magnitude higher estimate of $n_\textrm{b}$ by the latter \cite{Taskin2009} well outside the CNP region.
{Berry phase $\varphi_\textrm{B} = 2\pi \beta$ can be estimated from SdH oscillations using a semiclassical description \cite{BerryPhaseMikitik1999} $\Delta R_{\textrm{xx}} = A_{\textrm{SdH}} \textrm{cos}  [2 \pi (\frac{H_\textrm{F}}{H} + \frac{1}{2} + \beta )]$, where $A_{\textrm{SdH}}$ is the oscillation amplitude, $H_\textrm{F}$ is the frequency in $1/H$, and $\beta$ is the Berry factor. Near CNP the obtained Berry factor is $\beta = 0.5 \pm0.06$ as expected for the topological Dirac particles, while outside the CNP region a trivial value of $\beta \simeq 0$ is obtained (Fig.~2i)}.

\vspace{3mm}
\noindent\textbf{Nonequilibrium defect compensation at low irradiation dose.}
The stability of net carrier density crucially depends on the terminal irradiation dose.
First we illustrate that for low terminal electron doses ($\phi \lesssim 0.1~\textrm{C}\cdot \textrm{cm}^{-2}$) the resistivity is not stable as it evolves from the \textit{n}-region after irradiation through CNP back into the \textit{p}-region. The experiment was to simply change the dwelling time at room temperature (RT) to allow for vacancies to diffuse. The result is the reverse conversion (Fig.~3a) from metallic-like resistivity on the \textit{n}-side just after irradiation, through insulating resistivity at CNP (after nearly 250 hours at RT), and back to a weakly semiconducting-like resistivity on the \textit{p}-side. As the system evolves across CNP  a weak antilocalization quantum interference correction \cite{Glazman2012} to classical magnetoresistance (MR) emerges in the non-equilibrated conversion region (Fig.~3b), with a complex MR structure and nonlinear Hall resistivity (see Supplementary Fig.~6),
likely reflecting a presence of two carrier types. From isochronal annealing experiments {(Supplementary Fig.~7 and Supplementary Note 2)} we estimate energy barriers controlling defect migration to be $\sim 0.8~\textrm{eV}$ corresponding to $\sim 9,000~\textrm{K}$, in line with literature values for vacancy migration in solids \cite{Damask-Dienes1963}. We conclude that at low terminal doses while defect migration is sluggish, the equilibrium is not attained.

\vspace{3mm}
\noindent\textbf{Achieving stability at CNP with high irradiation dose.}
Next we demonstrate that stability can be controlled and optimized in a reverse conversion process by the annealing protocol when the terminal electron dose is sufficiently high. This is the key part of the material modification process by electron irradiation; without it surface transport via electron irradiation could not be obtained \cite{vdBeek2013}.
We demonstrate this for the samples exposed to the dose ten times higher, $\phi = 1~\textrm{C}\cdot \textrm{cm}^{-2}$. Fig.~4a shows temperature dependence of sheet resistance $R_{\textrm{xx}}$ and conductance $G_{\textrm{xx}}$ of Bi$_2$Te$_3$ tuned back to CNP through a thermal protocol shown in Fig.~ 4b, where it remained for months of testing {(Supplementary Figs.~ 8,9 and Supplementary Note 3)}.
As clearly seen in ARPES (Figs.~4c,d and Supplementary Fig.~10), under this annealing protocol Dirac dispersion remains preserved. The effect of tuning conduction from the irradiation induced \textit{n}-type (Fig.~4c) back toward CNP (Fig.~4d) is a relative shift of the Fermi level toward the Dirac point, consistent with transport data and with the calculated band structure\cite{Zhang-NatPhys09} of Bi$_2$Te$_3$ (see cartoons in Figs.~2f-h).

$R_{\textrm{xx}}(T)$ increases exponentially as the temperature decreases, turning below $\sim 200~K$ into variable range hopping {(Supplementary Fig.~11 and Supplementary Note 4)}. This bulk behavior is cut off at low temperatures when the contribution from the surfaces becomes comparable
and a temperature-independent surface transport with minimum conductance \cite{Fuhrer-gating2012} $G_{\textrm{xx}}^{\textrm{min}} \cong 20~e^2/h$ reveals its quantum nature. The 2D character of this region is witnessed by MR that depends only on the transverse component of magnetic field $H_\perp = H cos \theta$ {(Figs.~4e-h)}. Unlike the quadratic-in-field MR of a conventional metal, the large ($\gtrsim 10\%$) magnetoresistance at CNP (Fig.~4a) is found to be linear in field {(Figs.~4f,h)}.

The topological protection of the surface states near CNP can be tested in at least three ways. One, by breaking time reversal symmetry (TRS) which should gap out Dirac states.
Another, by the appearance of topological Berry phase \cite{Qi2011} of $\pi$ {(see Fig.~2i)}. And yet another, by detecting two-dimensional (2D) weak antilocalization (WAL) of Dirac particles traveling through time reversed paths \cite{Glazman2012} associated with this Berry phase.  Fig.~4a shows that applying TRS-breaking magnetic field indeed causes $R_{\textrm{xx}}(T)$ at the lowest temperatures to upturn, showing localizing behavior consistent with opening of the Dirac gap (see, e.g., ref. [\onlinecite{SmB6-Fisk2014}]).
From this we conclude that near CNP, even with the complex Bi$_2$Te$_3$ band structure, topological conduction channels dominate.

\vspace{3mm}
\noindent\textbf{Stable 2D quantum transport at CNP.}
As we approach CNP, weak antilocalization (WAL) quantum correction to classical conductivity emerges at low magnetic fields as a characteristic positive magnetoresistance cusp \cite{Glazman2012} (inset in Fig.~ 4b). In close proximity to CNP the corresponding cusp in negative magnetoconductance scales with the transverse field $H_\perp = H cos \theta$ {(Fig.~4e,g)}, confirming its 2D character.
In this case, the number $n_\textrm{Q}$ of quantum conduction channels contributing to WAL can be estimated from 2D localization theory \cite{HLN1980}
\begin{equation}
\Delta G(B)  \simeq  \alpha \frac{e^{2}}{2\pi ^{2}\hbar}f(\frac{B_{\phi }}{B}),
\end{equation}
where $\Delta G(B)$ is the low-field quantum correction to 2D magnetoconductance, coefficient $\alpha = n_\textrm{Q}/2$ equals to 1/2 for a single 2D channel, $f(x) \equiv ln x -\psi(1/2 + x)$, $\psi$  is the digamma function, and field $B_\phi=\frac{\hbar}{4el_{\phi}^{2}}$ is related to the dephasing length  $l_{\phi }$ of interfering electron paths. In Bi$_2$Te$_3$ at CNP the fit (in Fig.~4e) yields $\alpha \simeq 1.26\pm0.1$ corresponding to $n_\textrm{Q} \sim 2$, smaller than $G_{\textrm{xx}}^{\textrm{min}}$.
The obtained value of $\sim 2$ is quite remarkable since `universality' of $n_\textrm{Q}$ has been questioned \cite{KapitulnikWAL2013} given a likely formation of subsurface two-dimensional electron gas (2DEG) states of bulk origin. In Bi$_2$Se$_3$, where the Dirac point is expected to coincide with CNP, the fit similarly yields $\alpha \simeq 1.12\pm0.1$ {(Fig.~4g)}, corresponding to two 2D quantum channels we associate with two independent topological surfaces.

Finally, we remark that using thermal protocol illustrated in Fig.~4b we found that once CNP is reached the system remains there for months on cycling, which was the duration of our experiments. This robustness suggests that at higher electron doses vacancies are likely correlated; e.g. di-vacancies can form \cite{Damask-Dienes1963,Divacancy2010} and it is known that point-defect complexes can be stabilized \cite{BoisBeuneu1987} in systems with multiple sublattices.  With the choice of electron beam energy and terminal electron dose controlling the stability of pairs, and with a suitably designed thermal tuning to charge neutrality, the high-energy electron irradiation offers a path to large scale access to topological states.

\newpage
\noindent {\textbf{Methods}\\
\noindent
\textbf{Crystal growth and structural characterization.} Single crystals of  Bi$_2$Te$_3$, and Ca(0.09\%)-doped Bi$_2$Se$_3$ were grown by the standard Bridgman-Stockbarger method using a vertical temperature gradient pull.  X-ray diffraction of crystals was performed in a Panalytical diffractometer using Cu K$\alpha~ (\lambda = 1.5405{\AA})$ line from Philips high intensity ceramic sealed tube (3~kW) X-ray source with a Soller slit (0.04 rad) incident and diffracted beam optics. Transmission electron microscopy was performed at Evans Analytical Group.

\vspace{2mm}
\noindent
\textbf{Angularly resolved photoemission spectroscopy (ARPES).}
ARPES measurements were performed on the FemtoARPES setup \cite{LaserARPES-RSI2012} using a Ti:sapphire laser operated at 0.25 MHz repetition rate.
The specimens were probed with the 4th harmonic (at 6.26 eV), with an overall spectral resolution better than 30 meV. The data were taken along \textit{$\Gamma$-M} and \textit{$\Gamma$-K } directions. We found no substantial differences for measurements performed on several cleaved surfaces.

\vspace{2mm}
\noindent
\textbf{Electron irradiation.}
Electron irradiations were carried out in NEC Pelletron-type electrostatic accelerator in Laboratoire de Physique des Solides at \'{E}cole Polytechnique, Palaiseau, configured with a low temperature target maintained at $\sim 20$ K in a chamber filled with liquid hydrogen fed from a close cycle refrigerator.
All irradiations were performed with samples kept at 20 K, below the mobility threshold of the interstitials which tend to be more mobile than vacancies \cite{Damask-Dienes1963} -- this ensured the stability of all charges introduced by the irradiation process. It also allowed us to evaluate thermal migration of interstitials during the annealing process.
The size of the 2.5 MeV electron beam spot was reduced to $5~\textrm{mm}$ by a circular diaphragm aperture, with uniformity
of the beam current ensured with a constant beam sweep in \textit{x} and \textit{y}-directions at two non-commensurate frequencies. Careful calibration of the beam current density, typically of $2~\mu \textrm{A}$ on  a $0.2~\textrm{cm}^2$ surface, was performed by periodic introduction of a Faraday cage
between the diaphragm and
the sample chamber and the measurement of current collected on the control metallic sample. Beam current densities, limited to $10~\mu \textrm{C} \cdot \textrm{cm}^{-2}\textrm{s}^{-1}$ by the cooling rate, allowed modifications of carrier concentration on the order of $10^{20}$ \textrm{cm}$^{-3}$.

\vspace{2mm}
\noindent
\textbf{Transport measurements.}
The system was set up for \emph{in situ} transport measurements which could be monitored as a function of electron dose in real time. Crystals used in the irradiation experiments were typically $15~\mu \textrm{m}$ thick with spark-weld electrical contacts placed in van der Pauw contact configuration. \textit{Ex situ} transport and Shubnikov-de Haas oscillations measurements were performed in Quantum Design Physical Property Measurement System equipped with 14 T magnet. In \textit{ex situ} measurements, when samples were exfoliated to thicknesses of 200 nm and below, lithographically designed contacts with Ti/Au metallurgy were used (see Fig. 4a). Annealing experiments up to $200^\circ  \textrm{C}$ (such as shown in Fig. 4b and {Supplementary Note 2)} were performed in a box furnace in flowing nitrogen.


\small

\vspace{2mm}
\newpage
\noindent \textbf{Acknowledgements} We thank Boris Spivak for his insightful comments. We are grateful to
Martin Krusin and Vadim Oganesyan for the critical reading of the manuscript. This work was supported by the NSF grants DMR-1312483-MWN and DMR-1420634 (L.K.-E.), DOD-W911NF-13-1-0159 (L.K.-E.), ANR-13-IS04-0001-01 (M.K. and L.P.), EMIR project 11-11-0923 (M.K. and L.K.-E.), and NCN grant No. 2011/03/B/ST3/03362 (A.W.). The work of M.M. is supported by "Investissements d'Avenir" LabEx PALM (ANR-10-LABX-0039-PALM).

\vspace{1mm}

\noindent \textbf{Author contributions} Experiments were designed by L.K.-E. and M.K..  A.H. grew Bi$_2$Te$_3$ and Bi$_2$Se$_3$:Ca crystals and A.W. selected pristine crystals with low carrier densities. I.K. performed structural and chemical characterization of all crystals. I. K. and M. B. fabricated exfoliated structures. Transport measurements were performed by L.Z., H.D. and M.K., data analysis was done by L.Z., M.K. and L.K.-E. ARPES measurements were carried out by E.P., L.P. and M.M.. L.K.-E. wrote the manuscript with critical input from L.Z. and M.K.

\vspace{1mm}

\noindent \textbf{Additional information} The authors declare that they have no competing financial interests. Supplementary information accompanies this paper on www.nature.com. Correspondence and requests for materials should be addressed to L. K.-E.

\normalsize

\vspace{15mm}

\noindent\section*{FIGURE LEGENDS}
\vspace{-3mm}
\noindent \textbf{Figure 1 $\mid$ Tuning bulk conductivity of topological materials by swift particle irradiation.}
\textbf{a}, Energetic electron beams can penetrate solids to a depth of many tens of microns.
Electron irradiation affects the bulk but not the robust topological surfaces. \textbf{b}, Impinging electrons induce formation of Frenkel vacancy-interstitial pairs (inset), which act to compensate the intrinsic bulk defects. Main panel: Calculated crosssections $\sigma$ for Frenkel pair production in Bi, Te and Se sublattices as a function of electron energy $E$, assuming displacement energy \cite{Damask-Dienes1963} $\sim 25~ \textrm{eV}$. Energy thresholds, $E_{th}$, in $\sigma$ are set by atomic weight;
choosing $E <E_{\textrm{th}}^{Bi}$  or $E >E_{\textrm{th}}^{Bi}$ allows to tune Fermi level in both \textit{p-} and \textit{n-}type TIs. \textbf{c}, Transmission electron microscopy image of Bi$_2$Te$_3$ with electron dose $\phi = 1~\textrm{C}\cdot \textrm{cm}^{-2}$; the atomic displacements of $\sim1$ per 5,000 are not seen. {Scale bar is 1 nm}. \textbf{d}, ARPES spectrum taken at 130 K (see Methods) of a Bi$_2$Te$_3$ crystal after irradiation with high electron dose of $1.7~\textrm{C}\cdot \textrm{cm}^{-2}$. The irradiated sample becomes $n$-type, with Dirac point at $E_{DP} = -290(10)~\textrm{meV}$ relative to the Fermi level $E_F$ (just below the bottom of the bulk conduction band). It demonstrates electronic robustness of Dirac spectrum against electron irradiation. \textbf{e}, Resistivity of \textit{p-}type Bi$_2$Te$_3$ irradiated with 2.5~MeV electrons \textit{vs.} dose $\phi$ (red squares) measured \textit{in situ} at 20 K shows about three orders of magnitude increase at the charge neutrality point (CNP) where the conduction is converted from \textit{p-} to \textit{n-}type, moving the Fermi level $E_F$ across the Dirac point (see cartoon). Cycling to room temperature reverses the process, which can be recovered by further irradiation (blue circles) and stabilized. The conversion from \textit{n-} to \textit{p-}type is obtained in a material such as Pb-based TI, where vacancies on the Pb sites are known to be of acceptor type.

\vspace{3mm}

\noindent \textbf{Figure 2 $\mid$ Transport and quantum oscillations across charge neutrality point.} \textbf{a}, Top: Longitudinal resistivity $\rho_{\textrm{xx}}$ of Bi$_2$Te$_3$ crystals irradiated to different terminal doses and measured at 4.2 K shows three orders of magnitude increase at CNP where the bulk is converted from \textit{p- }to \textit{n-}type, as indicated by the sign change of the slope $dR_{\textrm{xy}}/dH$ of Hall resistance $R_{\textrm{xy}}$ shown in (\textbf{b}). Middle: Inverse Hall coefficient $1/ e R_\textrm{H} \simeq - {n_\textrm{b}}$ gives an estimate of net carrier density $n_\textrm{b}$ as a function of dose. At CNP $R_\textrm{H}$ changes sign. Bottom: Mobility just above CNP appears higher than just below, with the expected trend of $\mu$ at CNP indicated by dash.
Outside the immediate vicinity of CNP effective carrier mobilities $\mu$ estimated from $\rho_{\textrm{xx}} = \frac{1}{e n_\textrm{b} \mu}$ in the Drude model
are not much affected by the irradiation process, confirming that scattering events by Frenkel-pair point defects are scarce and the main effect is charge compensation.
Shubnikov-de Haas (SdH) quantum oscillations of $\Delta R_{\textrm{xx}}$ and $\Delta R_{\textrm{xy}}$ in Bi$_2$Te$_3$ with magnetic field applied along the \textit{c}-axis: \textbf{c}, before \textit{e}-beam exposure, \textbf{d}, irradiated to a dose $89~\textrm{C}\cdot \textrm{cm}^{-2}$ and \textbf{e}, to a dose $99~\textrm{C}\cdot \textrm{cm}^{-2}$. The corresponding sizes of Fermi surfaces obtained from SdH ( Supplementary Table 1) are cartooned as circles. \textbf{f}, \textbf{g}, \textbf{h}, Low-field $R_{\textrm{xy}}$ for the samples in \textbf{c}, \textbf{d}, and \textbf{e}. The Fermi level (dashed line) is moving from the bulk valence band (BVB) in (\textbf{f}) to the bulk conduction band (BCB) in (\textbf{h}) as illustrated in outsets by the cartoons of calculated \cite{Zhang-NatPhys09} Bi$_2$Te$_3$ band structure. \textbf{i}, The Landau level index plot \textit{vs.} field minima in the SdH oscillations in \textbf{c} (red) and \textbf{d} (blue) yields an estimate of Berry phase $\varphi_B = 2\pi \beta$, see text. \textbf{j,k}, Cartoons of a TI with conducting bulk before irradiation (\textbf{j}) and after, with ideally only topological surfaces conducting (\textbf{k}).

\vspace{3mm}

\noindent \textbf{Figure 3 $\mid$ Time evolution of charge transport across charge neutrality point in Bi$_2$Te$_3$ after `low-dose' exposure.}
\textbf{a}, Evolution of longitudinal resistivity $\rho_{\textrm{xx}}$ measured at 4.2 K after cycling to room temperature (RT) for a crystal irradiated with dose $\phi = 90~\textrm{C}\cdot \textrm{cm}^{-2}$.  Each RT dwell time is coded with a different color. Resistivity is seen to cross charge neutrality point (CNP) in reverse from  \textit{n-}type back to  \textit{p}-type. It is consistent with slow migration (hundreds of hours at RT) of vacancies (in accord with $\sim 0.8~eV$ migration barriers, see text) and shows that CNP can be reached by designing a suitable thermal protocol. Insets show $\rho_{\textrm{xx}}(T)$  for \textit{n-}type region (upper left) and \textit{p-}type region (lower right). \textbf{b}, Change in  magnetoconductance (MC) at different RT dwell times; here MC evolves from a quadratic field dependence of a typical bulk metal at short RT dwell times, through a complex region dominated by the charge-inhomogeneous bulk, to a weak antilocalization (WAL) region showing the characteristic low-field cusp near CNP. Here the data were normalized to the value at zero field. A fit to 2D localization theory \cite {HLN1980} is shown as red line, see text.

\vspace{3mm}

\noindent \textbf{Figure 4 $\mid$ Stable charge neutrality point and 2D conductance in e-beam irradiated topological materials.}
\textbf{a}, Sheet resistance $R_{\textrm{xx}}$ \textit{vs}. temperature of Bi$_2$Te$_3$ crystal at the charge neutrality point (red line) exhibits a plateau at low temperatures - a thumbprint of 2D surface conduction, see right panels in (\textbf{e}) and (\textbf{f}). We note that in Bi$_2$Te$_3$, CNP and Dirac point do not coincide, with the Dirac point situated within a valley in the bulk valence bands \cite{Zhang-NatPhys09}, and $G_{\textrm{xx}}^{\textrm{min}}$ is expected to reflect that \cite{Glazman2012}.
Magnetic field breaks time reversal symmetry and gaps Dirac bands, resulting in localizing behavior shown as dash. Inset: Optical image of the crystal showing van der Pauw contact configuration used. \textbf{b}, Annealing protocol with time steps $\Delta t = 30~ \textrm{min}$ implemented to tune Bi$_2$Te$_3$ crystal with dose $1~\textrm{C}\cdot \textrm{cm}^{-2}$ back to stable CNP. Inset: Magnetoresistance at 1.9 K after each annealing step, with colors matched to indicate different annealing temperatures.
\textbf{c,d}, ARPES spectra of a Bi$_2$Te$_3$ crystal irradiated with electron dose of $1.7~\textrm{C}\cdot \textrm{cm}^{-2}$ taken along $\Gamma - K$ direction in the Brillouin zone. \textbf{c}, Prior to annealing the irradiated sample is $n$-type, with Dirac point at $E_{DP} \sim -290(10)~\textrm{meV}$ relative to the Fermi level $E_F$. \textbf{d}, After annealing at $120^\circ$C Dirac point upshifts to a binding energy of $E_{\textrm{DP}} \sim -160(10)~\textrm{meV}$. The same shift is seen for the scans along $\Gamma - M$.
\textbf{e}, WAL low-field quantum interference correction to the linear-in-field magnetotransport (\textbf{f}, also see text) at CNP in a Bi$_2$Te$_3$ crystal at 1.9 K with its characteristic low-field cusp. The 2D character of WAL is evident in its scaling with transverse field $H_\perp = H cos \theta$, where $\theta$ is the tilt angle of the field measured from sample's \textit{c}-axis.
A fit to 2D localization (HLN) theory \cite{HLN1980} (solid line) confirms that the contribution is only from two surfaces and yields a dephasing field $B_{\phi }=\frac{\hbar}{4el_{\phi}^{2}} \sim 0.01~\textrm{T}$. \textbf{f}, Linear magnetoresistance at CNP shows 2D scaling with $H_\perp$.
\textbf{g}, WAL contribution at CNP in a Bi$_2$Se$_3$:Ca(0.09\%) crystal at 1.9 K also scales with $H_\perp$. At high fields outside the cusp, the scaling is seen to fail for $\theta \gtrsim 60^\circ$.
A fit to HLN theory (solid line) again confirms the contribution only from two surfaces and yields a smaller dephasing field $B_{\phi }\sim 0.004~\textrm{T}$ (corresponding to dephasing length $l_\phi \sim 220~\textrm{nm}$). \textbf{h},  Linear magnetoresistance at CNP also scales with $H_\perp$.

\newpage

\vspace{200mm}
\newpage
\eject

\begin{center}
\hspace{-20mm}
\includegraphics[width=18.7cm]{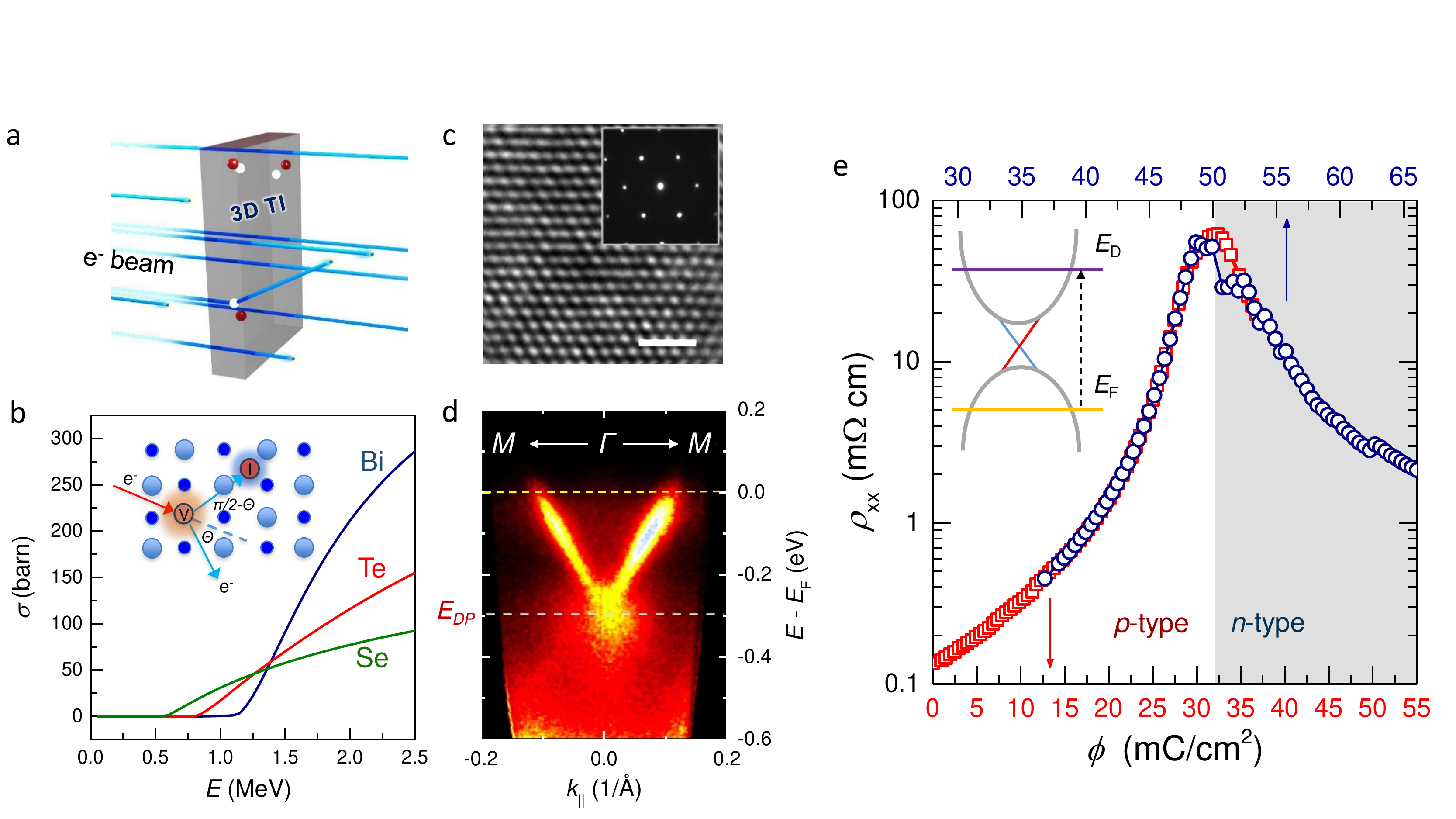}
\vfill\hfill Fig.~1 irrad; {ZZ \it et al.} \eject

\hspace{-50mm}
\includegraphics[width=29cm]{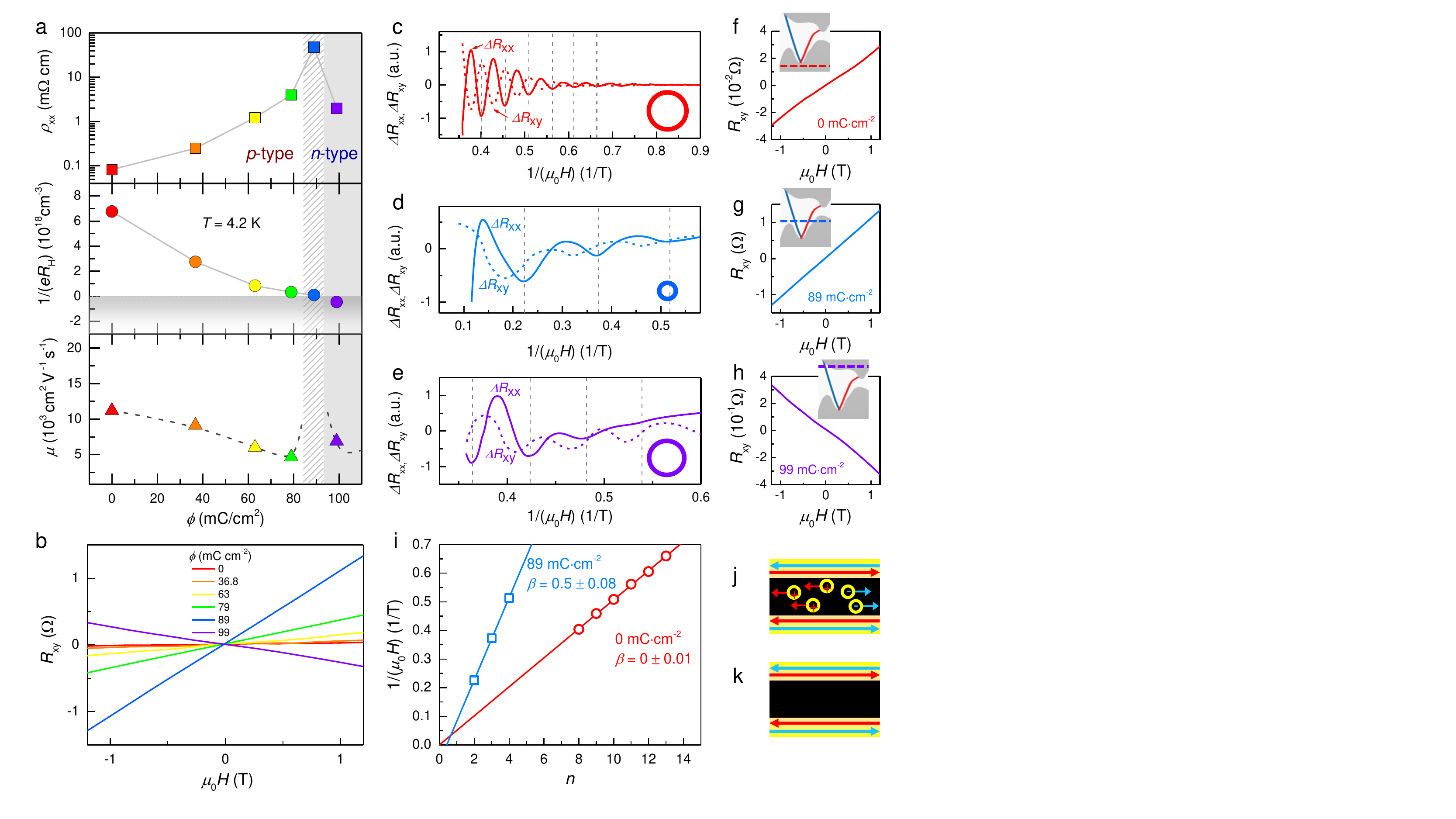}
\vfill\hfill Fig.~2 irrad; {ZZ \it et al.} \eject

\hspace{-40mm}
\includegraphics[width=24cm]{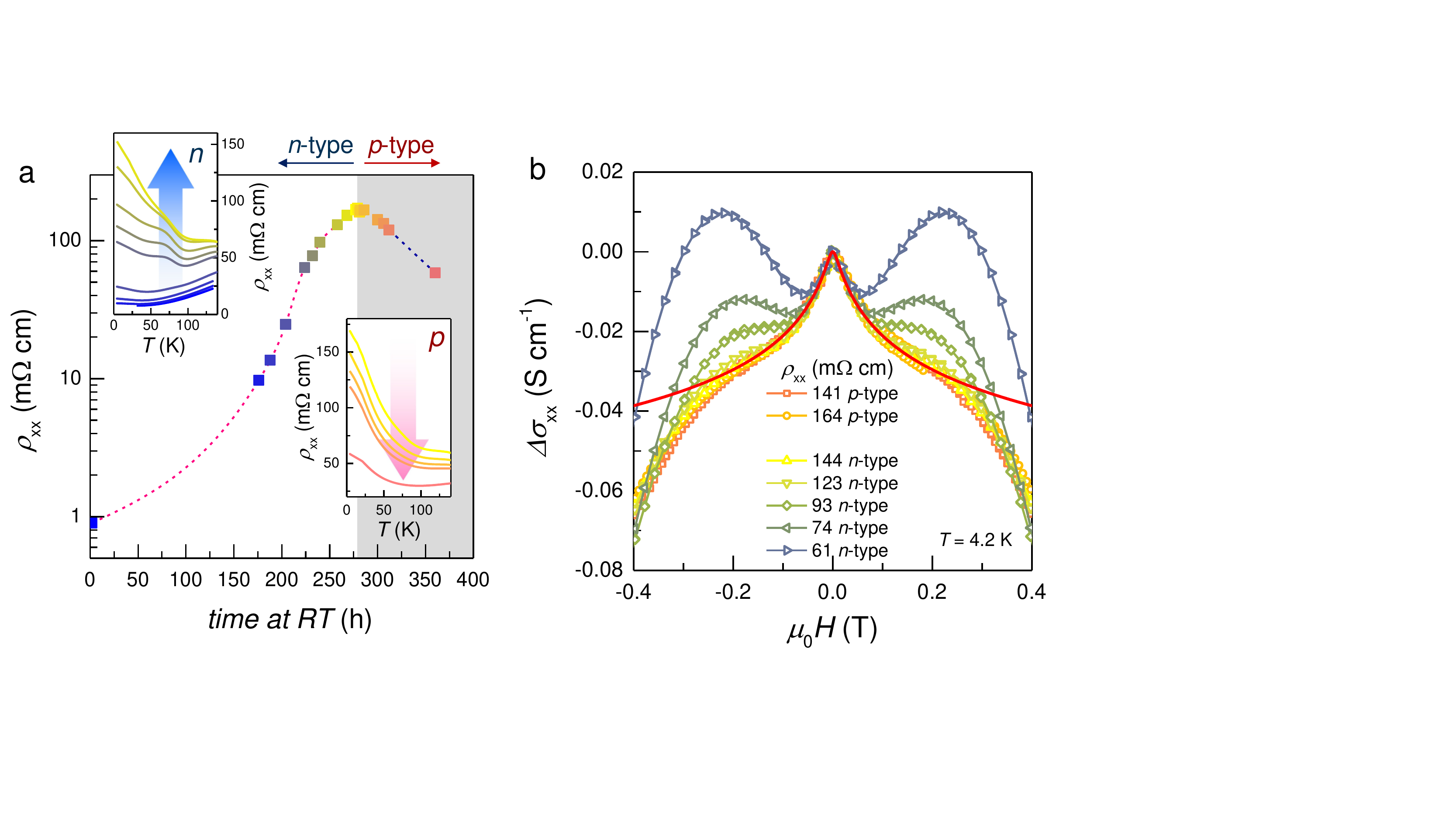}
\vfill\hfill Fig.~3 irrad; {ZZ \it et al.} \eject

\hspace{-28mm}
\includegraphics[width=19.0cm]{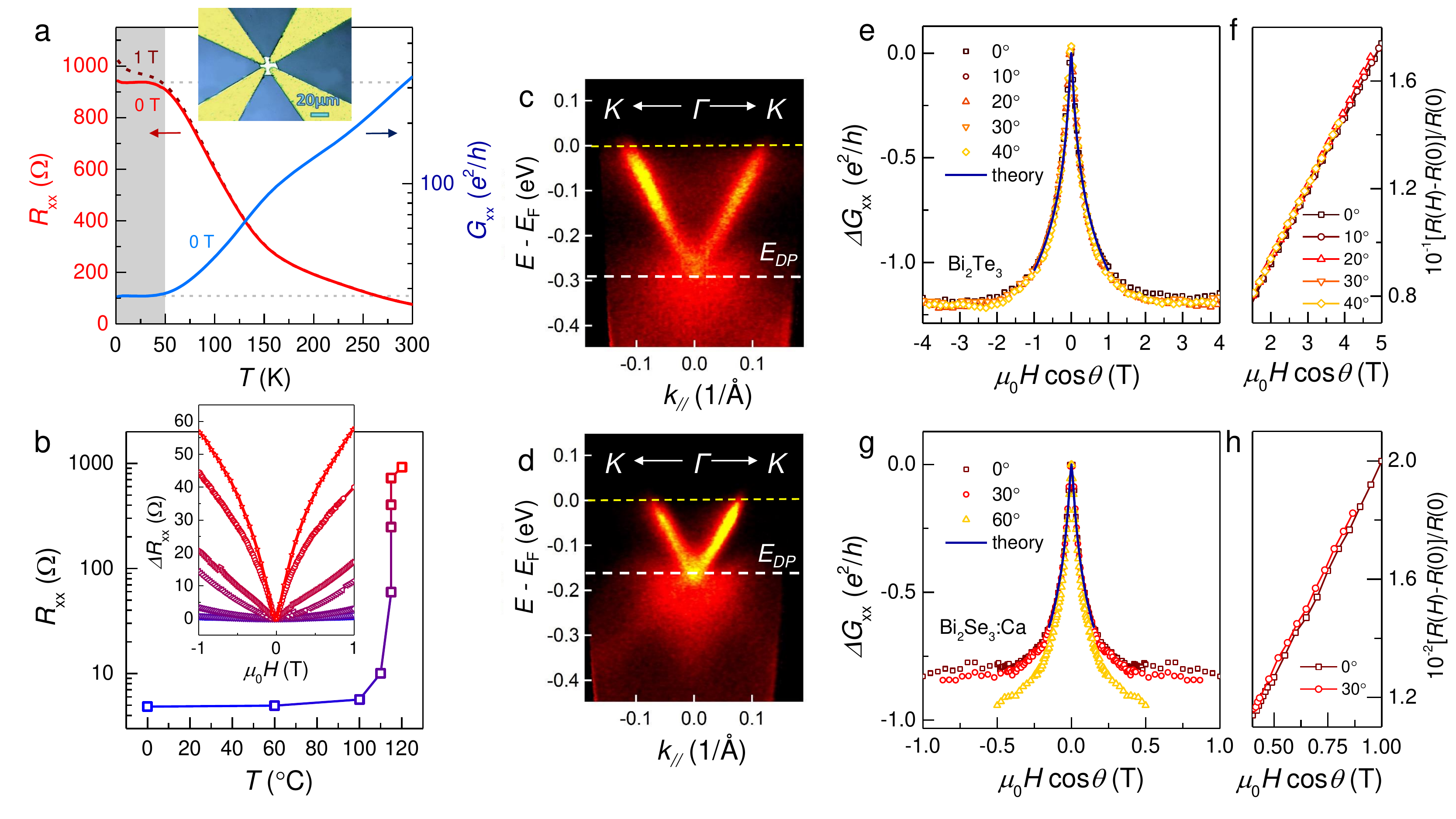}
\vfill\hfill Fig.~4 irrad; {ZZ \it et al.} \eject

\end{center}
\end{document}


\noindent\textbf{\large{\underline{Supplementary Figures}}}\\
\begin{figure}[H]
\vspace{-10mm}
\begin{center}
\includegraphics[width=\linewidth]{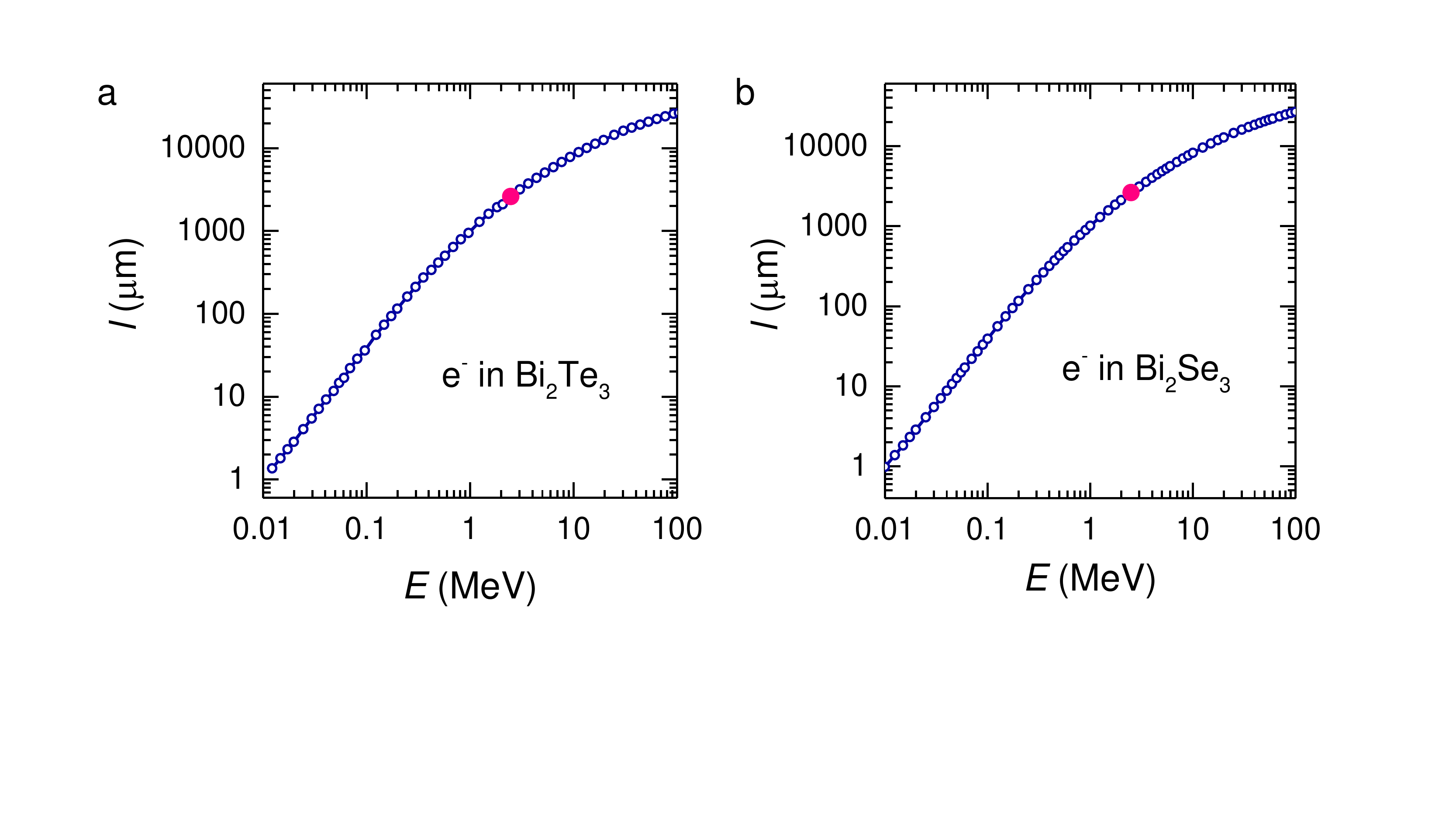}
\vspace{-30mm}
\end{center}
\protect
\end{figure}

\textbf{Supplementary Figure 1$\mid$ Penetration range of electrons beams in Bi$_2$Te$_3$ and Bi$_2$Se$_3$.}
Penetration depth of electrons in (a) Bi$_2$Te$_3$ and (b) Bi$_2$Se$_3$ calculated using NIST ESTAR simulator (http://physics.nist.gov/PhysRefData/Star/Text/ESTAR.html). The penetration depth at 2.5 MeV (red dots) used in our experiments is $ > 2,000~ \mu \textrm{m}$ and the resulting depth profile of vacancies is uniform over hundreds of microns.

Particle beams such as Ne$^+$ and He$^{2+}$ (alpha particles) have a much shorter penetration depth than electrons and are known to produce defect cascades and extended defects. For example, 150 keV  Ne$^+$ ion irradiation has a nonuniform depth profile and shows a pileup of defects at about $0.2\mu \textrm{m}$.
Such irradiations were shown to \textit{increase} electron concentration away from the surface states and to stabilize Fermi level high up in the bulk conduction bands \cite{Furdyna2014} of \textit{n}-type Bi$_2$Te$_3$ and Bi$_2$Se$_3$. We  note that charge doping by electron irradiation in band semiconductors such as PbTe has been explored in the eighties \cite{MarcinK1989}. One previous attempt in Bi$_2$Te$_3$ \cite{vdBeek2013} used low \textit{e}-beam fluences with the target kept at room temperature. In this process, the irradiated samples displayed none of the features associated with surface carriers or even showed semiconducting like behavior of resistivity \textit{vs}. temperature. In our experiments, irradiation were performed in liquid hydrogen. This allowed us to control pair formation during the energetic irradiation process and develop a scheme under which stable vacancy content that compensated native charged defects was established.
\newpage

\begin{figure}[H]
\vspace{-10mm}
\begin{center}
\includegraphics[width=17cm]{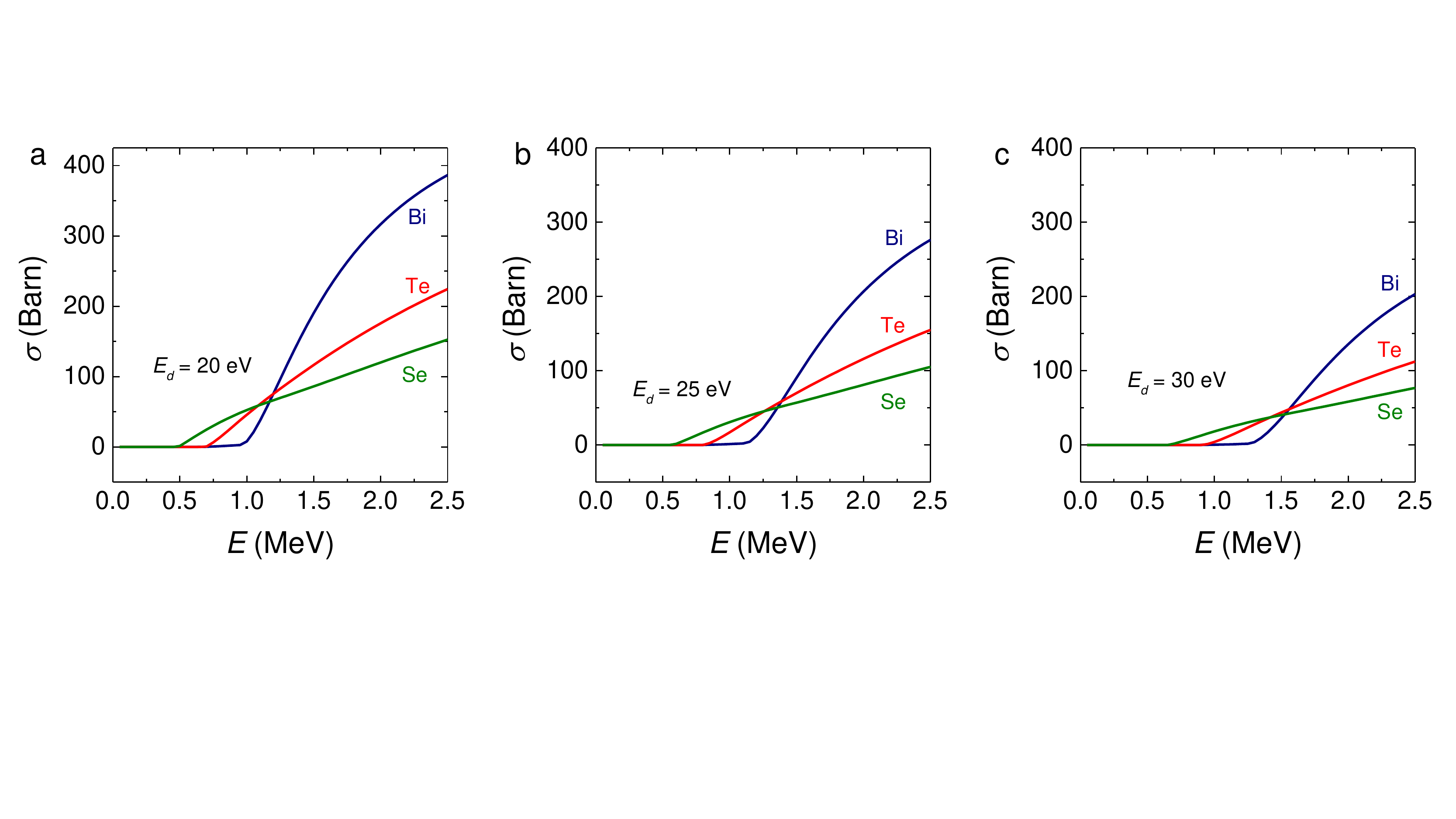}
\vspace{-35mm}
\end{center}
\protect
\end{figure}

\noindent \textbf{Supplementary Figure 2$\mid$ Energy thresholds for the Frenkel pair creation on Bi, Te and Se sublattices.}
Calculated crosssections $\sigma$ for Frenkel pair production on Bi, Te and Se sublattices as a function of electron energy $E$ with the displacement energy $E_\textrm{d}$ of (a) 20 eV, (b) 25 eV, and (c) 30~eV. Energy thresholds, $E_{\textrm{th}}$, in $\sigma$ are not sensitive to $E_\textrm{d}$ and are largely set by the atomic weight.

Empirical determination of displacement energies $E_\textrm{d}$ is relatively straightforward in metals \cite{Damask-Dienes1963}.  It is much more complex in semiconductors or semimetals where defects can affect both carrier concentration and mobility and there is an argument given by Seitz \cite{Seitz1949} that $E_\textrm{d}$ can be related to sublimation energy. Damask-Dienes \cite{Damask-Dienes1963} give $E_\textrm{d}$ for Bi, Te and Se in the range of 20 - 30 eV. We have performed calculations of Frenkel-pair production crosssections for all three sublattices in this range. The calculation is based on Rutherford collision of a relativistic electron with energy $E$ with a nucleus of mass $M$, where the transferred energy $E = E_{\textrm{max}} \textrm{sin}^2(\theta)/2$  may range from zero in a glancing collision to a maximum energy $E_{\textrm{max}} = 2E(E+2mc^2)/Mc^2$ transferred in a head-on collision. The pair production crosssection is given by $\sigma = \frac{2\pi}{E_{\textrm{max}}} \int_{E_\textrm{d}}^{E_{\textrm{max}}} \frac{\textrm{d}\sigma(E)}{\textrm{d}\theta} P_\textrm{d}(E)\textrm{d}E$, where the differential crosssection is a relativistic extension of classical Rutherford formula derived by Mott \cite{Mott1929,Mott1932} and $P_\textrm{d}(E)$ is the probability the atom was pushed in the direction $\frac{\pi}{2} - \frac{\theta}{2}$ (see main Fig.~1b).
Both the energy thresholds and the ratios of pair production on Bi and Te or Bi and Se sublattices are not very sensitive to $E_\textrm{d}$. For this reason we have chosen $E_\textrm{d}$  = 25 eV, in the midrange.

\begin{figure}[H]
\begin{center}
\includegraphics[width=\linewidth]{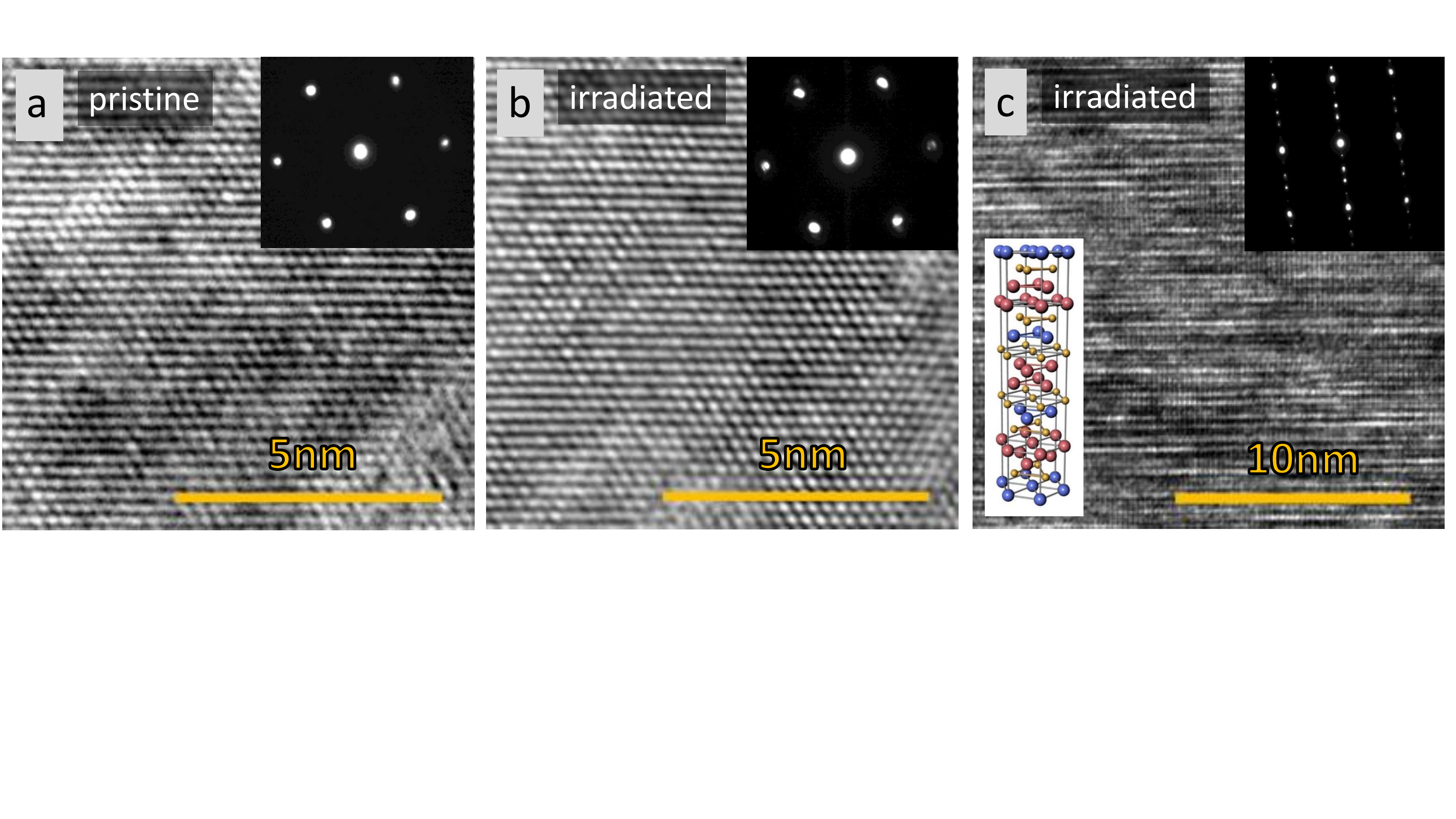}
\vspace{-40mm}
\end{center}
\protect
\end{figure}

\noindent \textbf{Supplementary Figure 3$\mid$ Transmission Electron Microscopy of Bi$_2$Te$_3$ exposed to 2.5 MeV e-beams.}
TEM images and diffraction spots of (a) pristine and (b) irradiated Bi$_2$Te$_3$ show the same hexagonal lattice in the $ab$-plane. (c) Layered van der Walls structure along the $c$-axis (normal to (00\={1}) cleavage plane) after irradiation. Inset: Rhombohedral layered structure of Bi$_2$Te$_3$ constructed using lattice parameters ($a = 4.38\AA$ and $c = 30.45\AA$) from the X-ray diffraction (XRD). The van der Walls structure has three quantuple layers per unit cell \cite{Qi2011}.

Shown are transmission Electron Microscopy (TEM) images of Bi$_2$Te$_3$ irradiated to a dose $\phi = 1 \textrm{C/cm}^2$. At this dose and after 40 hours at room temperature, we estimate the number of the atomic displacements as follows. For 2.5 MeV electrons the effective cross-section is $\sigma \sim 300$ barns (1 barn $= 1 \times 10^{-24}~ \textrm{cm}^2$). With the dose of $0.1~ \textrm{C/ cm}^2 = 6.25 \times 10^{17}$ electrons/cm$^2$ we have 0.000187 d.p.a on Bi sublattice, corresponding to a roughly $\sim1$ per 5,000 atoms ejected, which are not easily discerned in the image. The images of irradiated and pristine samples are indistinguishable.

We note that electron irradiation induced vacancy - interstitial (Frenkel) pairs \cite{Frenkel-pairsMurray1987,BoisBeuneu1988} can be well controlled by the integrated irradiation dose (beam fluence) to a very low carrier concentration level ($ \lesssim 10^{13}~ \textrm{cm}^{-3}$). Electron irradiation produces donor-type doping in both Bi$_2$Te$_3$ and Ca-doped Bi$_2$Se$_3$ compounds (with a slightly higher rate $\frac{\partial n_\textrm{b}}{\partial \phi}$ for Bi$_2$Se$_3$). At the present stage we cannot differentiate between the doping action of vacancies on Bi and Te/Se sites, however, we clearly demonstrate bulk charge compensation of \textit{p}-type TIs across charge neutrality point.

\begin{figure}[H]
\vspace{-8mm}
\begin{center}
\includegraphics[width=17cm]{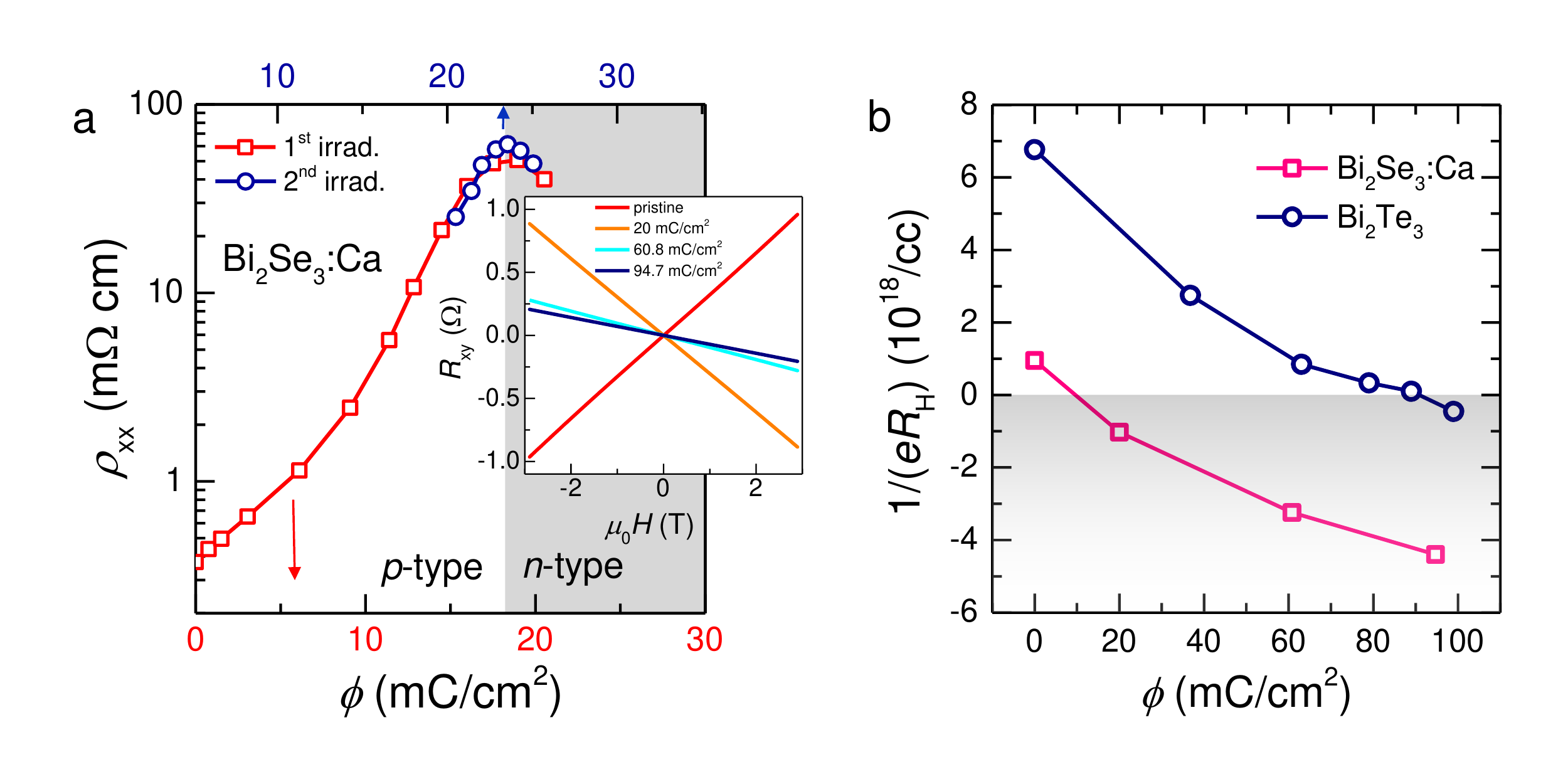}
\vspace{-12mm}
\end{center}
\protect
\end{figure}

\noindent\textbf{Supplementary Figure 4$\mid$ Tuning through charge neutrality point (CNP): \textit{p}- to \textit{n}- and \textit{n}- to \textit{p}- conversions.}
(a) Longitudinal resistivity $\rho_{xx}$ of the initially \textit{p-}type Bi$_2$Se$_3$ doped with 0.09\% Ca measured \textit{in situ} at 20 K in the irradiation chamber \textit{vs.} irradiation dose $\phi$. Initial ($1^{\textrm{st}}$) irradiation is shown as red squares.
The conversion of conductance from \textit{p-} to \textit{n}-type tales place at $\phi_{\textrm{max}} \cong 18~\textrm{mC/cm}^2$. Warming up to room temperature partially reverses the compensation process, which can be fully recovered with repeated irradiation (blue circles). The accumulated dose on the $2^{\textrm{nd}}$ irradiation at $\rho_{\textrm{xx}}^{\textrm{max}}$ is  $\phi_{\textrm{max}} \cong 24~\textrm{mC/cm}^2$. Inset: The change of sign of the slope ${\partial R_{\textrm{xy}}}/{ \partial H}$ of Hall resistance $R_{\textrm{xy}}$ from below $\phi_{\textrm{max}}$ to above.  (b) Inverse Hall coefficient $1/ e R_\textrm{H} \simeq - {n_\textrm{b}}$ gives an estimate of net carrier density $n_\textrm{b}$ as a function of dose. A quasilinear behavior of $1/ e R_\textrm{H}$ seen in both Bi$_2$Te$_3$ and Bi$_2$Se$_3$:Ca crystals. This universal rate of charge compensation can be used to establish the value of $\phi_{\textrm{max}}$ at the charge neutrality point in both, bulk crystals and thin films. The key starting point is the initial carrier concentration which determines the irradiation time. Our preliminary results on thin films on SiO$_2$ substrates confirm this.

\clearpage
\begin{figure}[H]
\vspace{-15mm}
\begin{center}
\includegraphics[width=17cm]{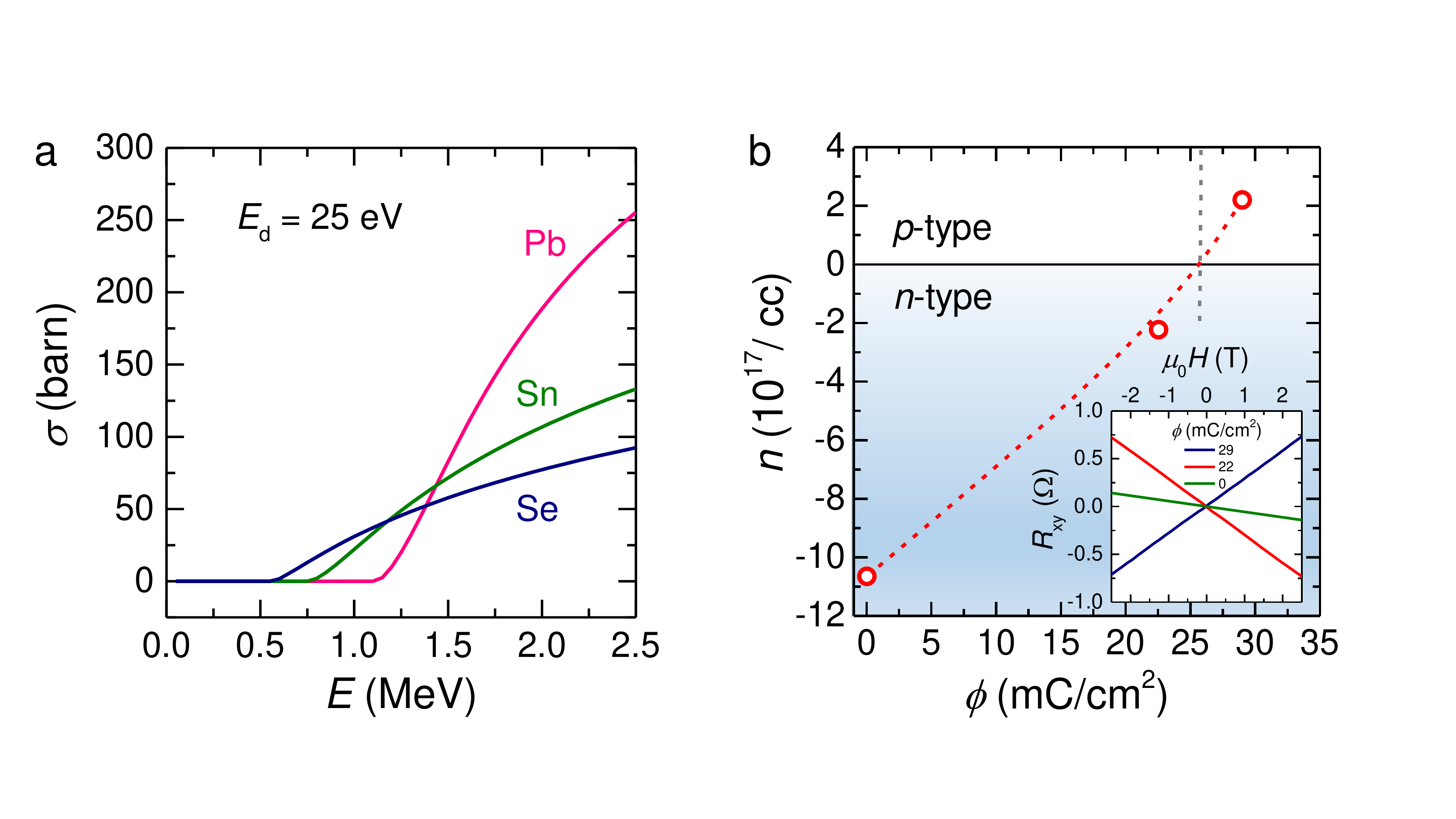}
\vspace{-15mm}
\end{center}
\protect
\end{figure}

\noindent\textbf{Supplementary Figure 5$\mid$ Calculated crosssections $\sigma$ for Frenkel pair production on Pb, Sn and Se sublattices.}
(a) Calculated crosssections $\sigma$ for Frenkel pair production on Pb, Sn and Se sublattices as a function of electron energy $E$ with the displacement energy $E_d = 25~\textrm{eV}$. (b) Carrier concentration $n_b$ at 4 K as a function of electron dose in a topological insulator Pb$_{1-x}$Sn$_x$Se ($x = 0.3$) irradiated at 20 K. It demonstrates conversion from the initially electron (\textit{n}-type) to hole (\textit{p}-type) conduction at a dose $\phi \cong 25~\textrm{mC/cm}^2$.

Electron irradiation can be used for both \textit{p}- to \textit{n}- and \textit{n}- to \textit{p}- conversions. Since the conversion is controlled by the dominant donor or acceptor character of (mainly) vacancy partners in Frenkel pairs, it can be tuned either by electron beam energy (which chooses the dominant sublattice) or (for a given beam energy) by a suitable choice of a sublattice in another system. An example of the latter is a Pb-based topological insulator \cite{PbSnSe-Story2012} Pb$_{1-x}$Sn$_x$Se, where preliminary electron irradiation experiments \cite{PbSnSe-Marcin2015} indicate the conversion of conduction from \textit{n}- to \textit{p}-type. This is consistent with the well known acceptor flavor of vacancies formed on the Pb sublattice \cite{Pb-vacancy1969}.

\clearpage

\begin{figure}[H]
\vspace{-5mm}
\begin{center}
\includegraphics[width=13cm]{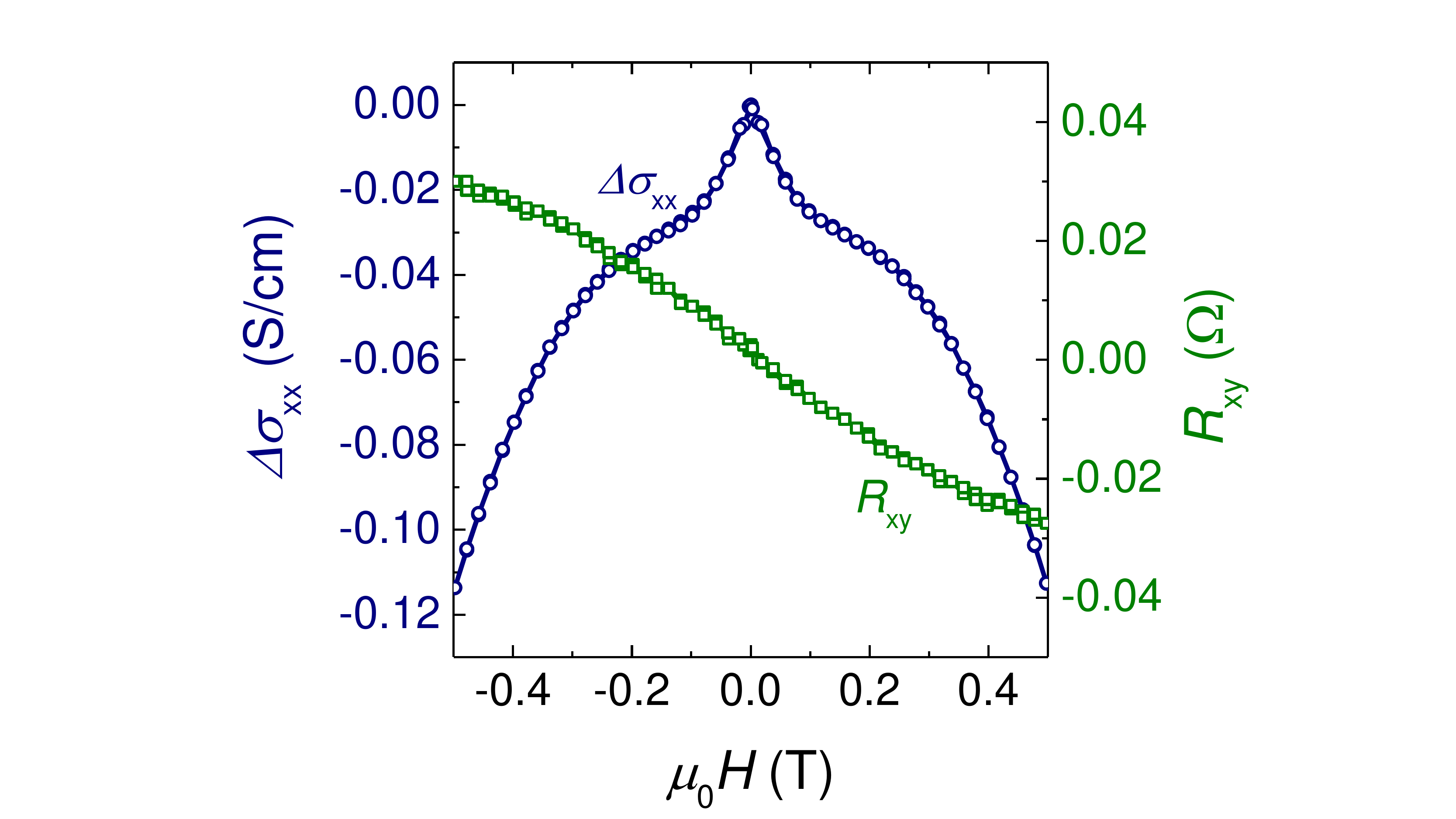}
\end{center}
\protect
\end{figure}

\noindent\textbf{Supplementary Figure 6$\mid$ Nonlinear Hall resistivity in the non-equilibrated state of irradiated Bi$_2$Te$_3$.} WAL cusp in magnetoconductance of Bi$_2$Te$_3$ crystal irradiated with a low electron dose of $\phi = 90~\textrm{mC/cm}{^2}$ after dwelling for 268 hours at room temperature (for full time evolution see main Fig. 3). It is riding on a large parabolic background typical of a bulk metal or semiconductor. The Hall resistance $R_{xy}$ shows nonlinearity which can evolve with time.

A complex temporal behavior of magnetoconductance $\sigma(H)$  is observed during room temperature annealing after electron irradiation with a relatively low dose. In this case the system is out of equilibrium and charge compensation can be locally incomplete, with two types of carriers likely to be present. In such a case Hall resistivity can be nonlinear, and this is indeed what we observe. It has been shown \cite{KapitulnikWAL2013} that a net WAL feature can arise from the quantized bulk channels which may also contribute the opposing weak localization correction, WL. The background from the bulk is time-dependent; $\sigma (H)$  changes slope from positive to negative at fields beyond the field range of WAL cusp. This background is evolving in time in a manner suggestive of spatial inhomogeneity of charge compensation.

\clearpage
\begin{figure}[H]
\vspace{-2mm}
\begin{center}
\includegraphics[width=16cm]{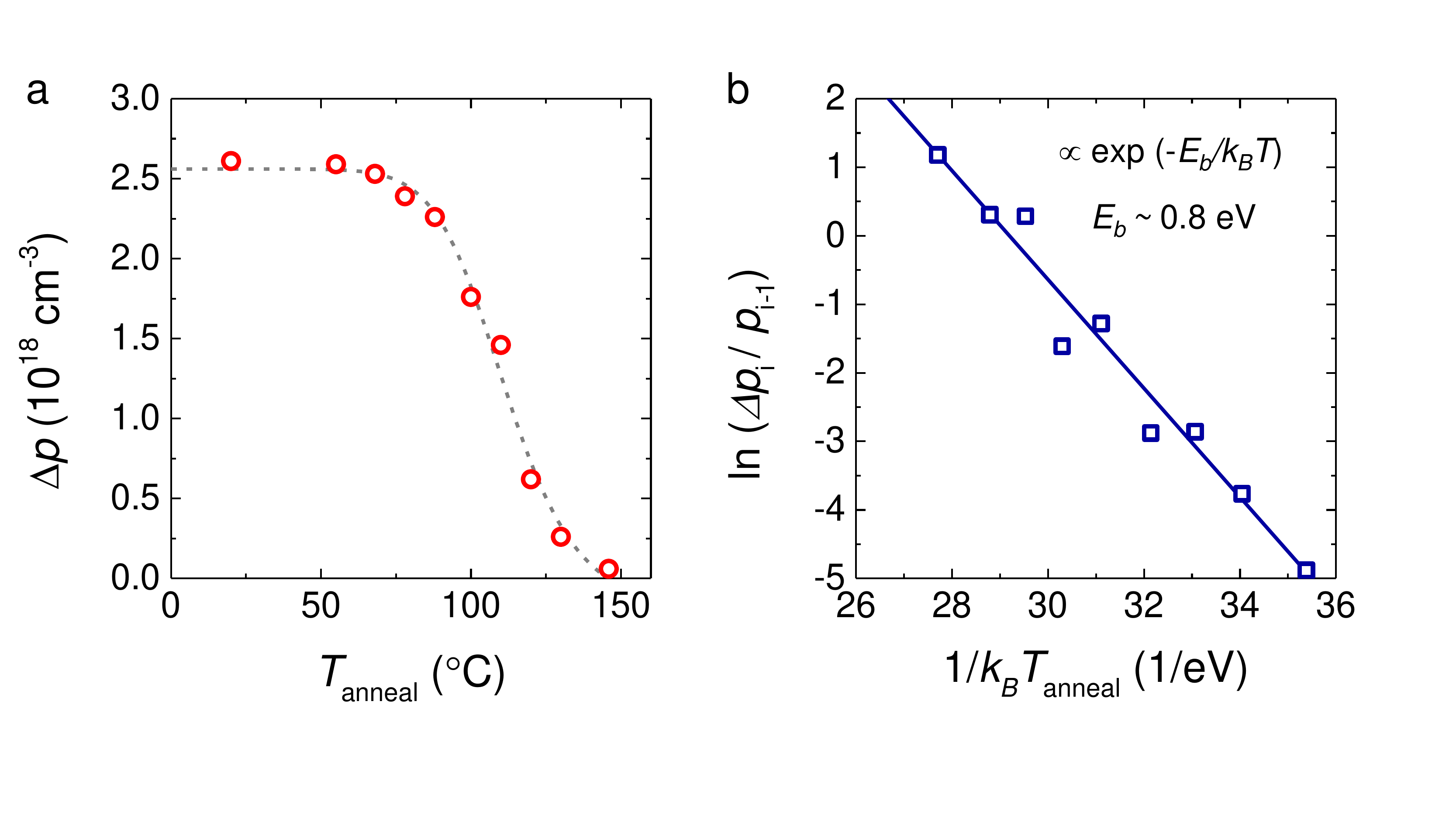}
\end{center}
\vspace{-20mm}
\protect
\end{figure}

\noindent\textbf{Supplementary Figure 7$\mid$ Isochronal annealing experiments and energy barriers for defect migration.} (a) Isochronal annealing of Bi$_2$Te$_3$ crystal irradiated to the dose of $76~ \textrm{mC/cm}^2$ in 30min intervals followed by the resistivity measurement at 4.2 K. (b)   Determination of the energy barrier to defect migration from the isochronal annealing step around $100^\circ \textrm{C}$.

\begin{figure}[H]
\vspace{-5mm}
\hspace{-5mm}
\begin{center}
\includegraphics[width=15cm]{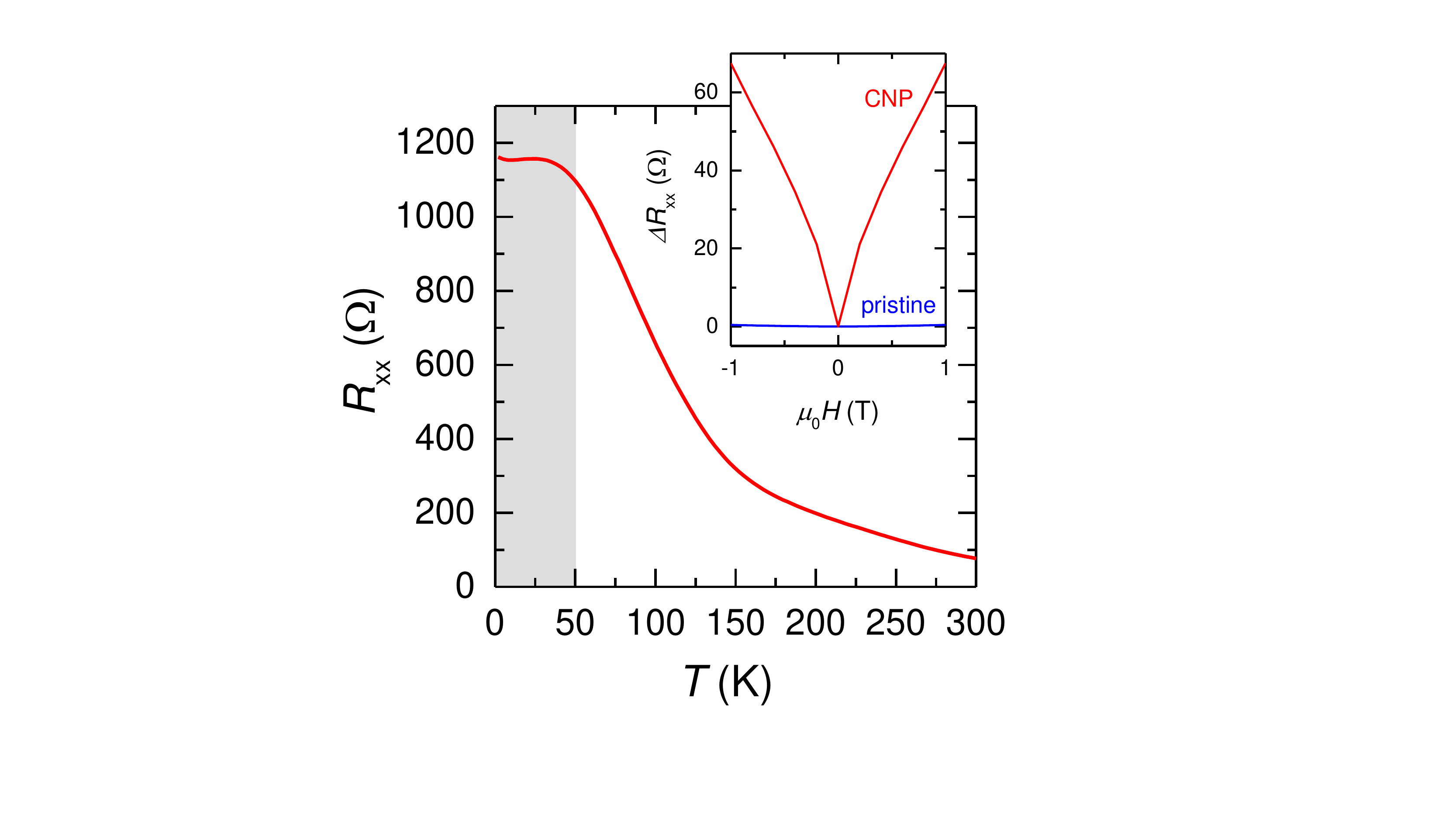}
\vspace{-20mm}
\end{center}
\protect
\end{figure}

\noindent\textbf{Supplementary Figure 8$\mid$ Longitudinal resistance at CNP of another Bi$_2$Te$_3$ crystal.} Sheet resistance $R_{\textrm{xx}}$ \textit{vs}. temperature of another Bi$_2$Te$_3$ crystal at the charge neutrality point reached through the annealing schedule shown in main Fig.~4b. Inset: Magnetoresistance at 1.9 K in a pristine Bi$_2$Te$_3$ crystal (blue) and of a crystal irradiated with electron dose of 1 C/cm$^2$ after annealing to CNP (red). At CNP a sharp weak antilocalization (WAL) cusp appears, which is not seen in either pristine crystals or after irradiation.

\clearpage
\begin{figure}[H]
\vspace{-15mm}
\hspace{-5mm}
\begin{center}
\includegraphics[width=15cm]{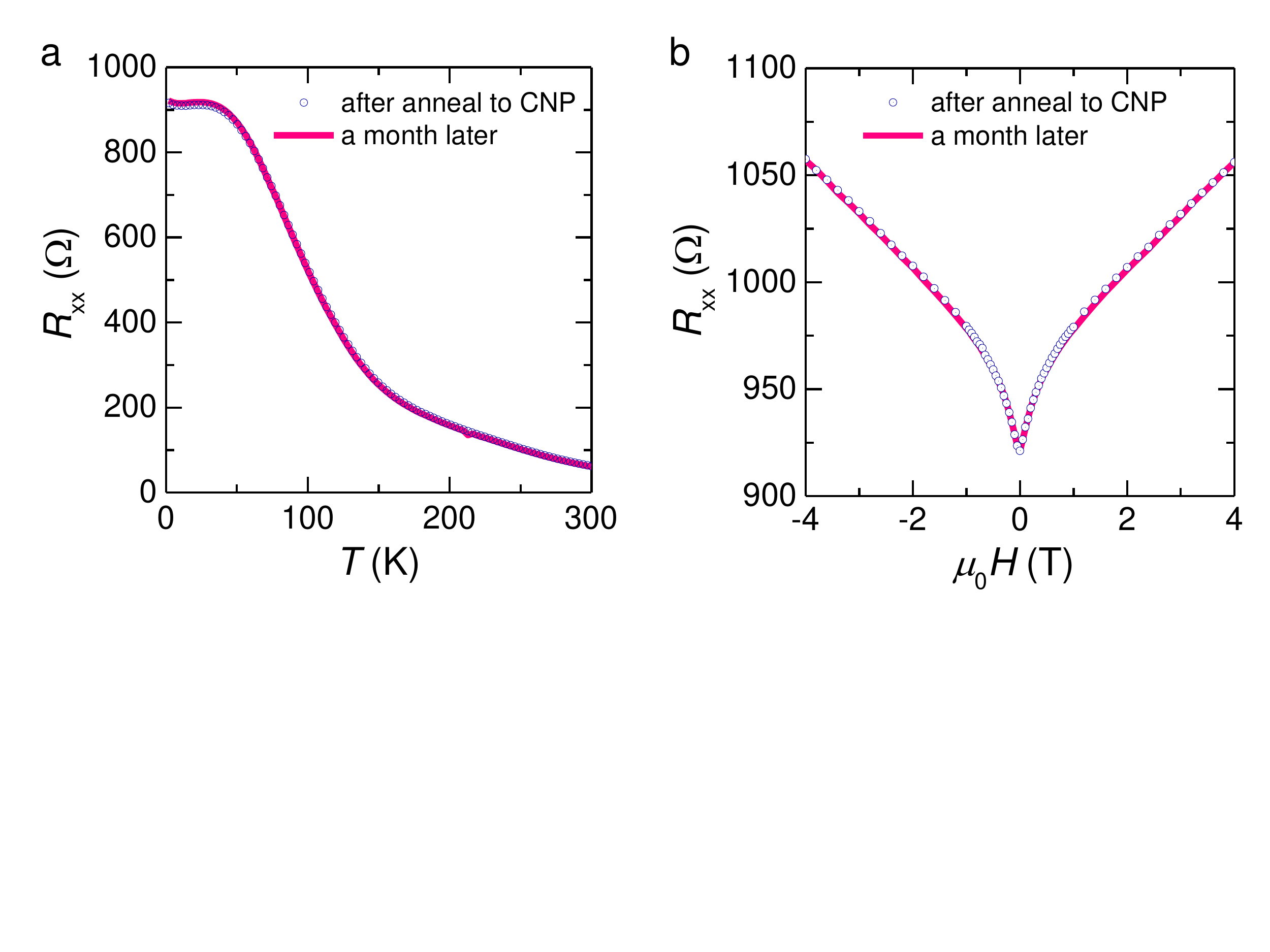}
\end{center}
\vspace{-50mm}
\protect
\end{figure}

\noindent\textbf{Supplementary Figure 9$\mid$ Stable charge neutrality point in Bi$_2$Te$_3$ achieved through annealing.} Sheet resistance $R_{xx}$ of the Bi$_2$Te$_3$ crystal in main Fig. 4 at the charge neutrality point (a) \textit{vs}. temperature and (b) vs. magnetic field showing the long-term temporal stability of the signal after high irradiation dose and anneal back to CNP.

\begin{figure}[H]
\vspace{-10mm}
\hspace{-20mm}
\begin{center}
\includegraphics[width=17cm]{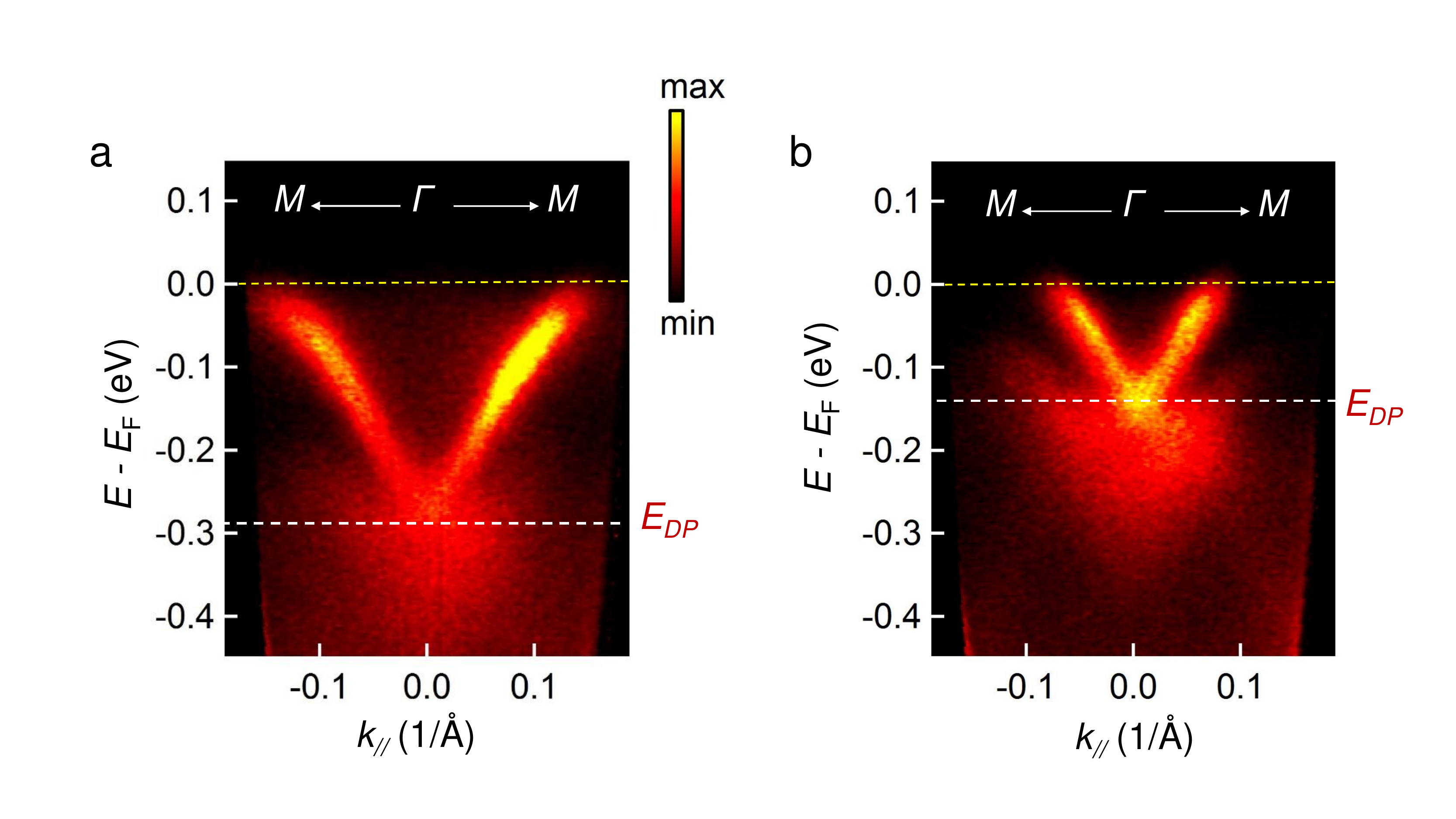}
\vspace{-20mm}
\end{center}
\protect
\end{figure}

\noindent\textbf{Supplementary Figure 10$\mid$ ARPES of irradiated and annealed Bi$_2$Te$_3$ crystals.} ARPES spectra of a Bi$_2$Te$_3$ crystal irradiated with electron dose of $1.7~\textrm{C/cm}^2$ taken along $\Gamma - M$ direction in the Brillouin zone. (a) Before annealing the Dirac point is at $E_{DP} \sim -290(10)~\textrm{meV}$ relative to the Fermi level $E_F$. (b), After 30 min anneal at $120^\circ$C Dirac point upshifts to  $E_{DP} \sim -160(10)~\textrm{meV}$.
ARPES data demonstrate that the annealing protocol that tunes the system back to CNP preserves Dirac states.

\clearpage
\begin{figure}[H]
\vspace{-20mm}
\hspace{-5mm}
\begin{center}
\includegraphics[width=16cm]{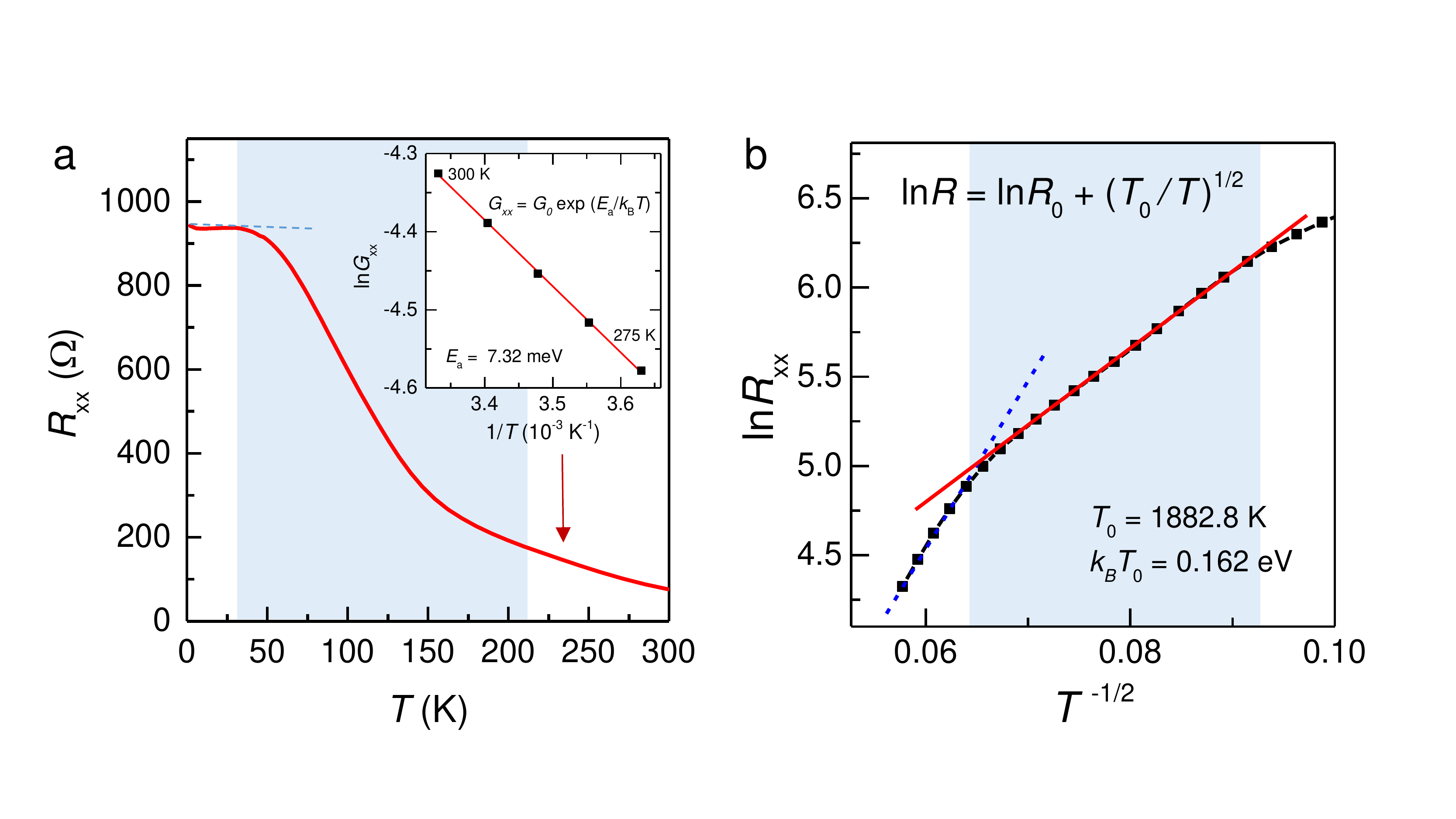}
\end{center}
\protect
\end{figure}

\noindent\textbf{Supplementary Figure 11$\mid$ Variable range hopping bulk charge transport at low carrier density.} a) Sheet resistance $R_{\textrm{xx}}$ \textit{vs}. temperature of Bi$_2$Te$_3$ crystal at the charge neutrality point. Inset: A fit to a simple activation law near room temperature. (b) A fit to Efros-Shklovskii variable range hopping (VRH) law \cite{Shklovskii2012} law in the temperature region marked in light blue.

\clearpage
\vspace{100mm}
\textbf{\large{\underline{Supplementary Tables}}}\\

\begin{figure}[H]
\begin{center}
\includegraphics[width=24cm]{TableS1d2}
\end{center}
\protect
\end{figure}
\vspace{-100mm}
\noindent \textbf{Supplementary Table 1$\mid$ Transport parameters for a pristine and irradiated Bi$_2$Te$_3$.}
1st column shows irradiation doses $\phi$, Fermi wavevectors $k_F$ and the corresponding carrier densities $n$ obtained from Shubnikov-de Haas (SdH) oscillations are in the 2nd and 3rd columns respectively. The inverse Hall coefficient which gives an estimate of carrier density at low magnetic fields is in the 4th column. The last two columns show carrier mobilities $\mu$ and mean free paths $\textsl{l}$.

\clearpage

\noindent\textbf{\large{\underline{Supplementary Notes}}}\\

\noindent \textbf{Supplementary Note 1\\
\noindent Tuning transport across CNP in Bi$_2$Se$_3$:Ca.}\\
Longitudinal resistivity $\rho_{\textrm{xx}}$ of the initially \textit{p-}type Ca-doped Bi$_2$Se$_3$ \cite{BridgmanBiSeCa2014,Ca-dopingOng2009}
measured \textit{in situ} in the irradiation chamber kept at 20 K as a function of 2.5 MeV electron irradiation dose shows behavior identical to Bi$_2$Te$_3$ under the same irradiation conditions (see main Fig.~1e). Supplementary Fig.~4 shows two orders of magnitude resistivity increase to a maximum $\rho_{\textrm{xx}}^{\textrm{max}}$ where the conversion from \textit{p-} to \textit{n-}type takes place defining charge neutrality point (CNP). Irradiation doses to achieve CNP are lower in Bi$_2$Se$_3$:Ca than in Bi$_2$Te$_3$ since the initial bulk carrier concentration is lower. As in Bi$_2$Te$_3$, for low irradiation doses $\rho_{\textrm{xx}}(\phi)$ is reversible upon temperature cycling to room and back to 20 K, with $\rho_{\textrm{xx}}(\phi)$ and $\rho_{\textrm{xx}}^{\textrm{max}}$ traced exactly by the next irradiation cycle. Our experiments on Bi$_2$Te$_3$ were carried on more than a dozen samples cut from the same large \textit{p}-type crystal and the rates of variation of carrier concentration \textit{vs}. dose in the range 10 mC/cm$^2$ to 1.6 C/cm$^2$ were essentially the same in all samples. In all measured Bi$_2$Se$_3$:Ca crystals the rates were consistently higher than in Bi$_2$Te$_3$.
We should note that access to Dirac point in Bi$_2$Se$_3$ is expected, in principle, to be less complicated than in Bi$_2$Te$_3$, since in Bi$_2$Se$_3$ DP is positioned in the bulk bandgap, while the band structure of Bi$_2$Te$_3$ (where Dirac point, DP, is nestled in the valley of the bulk valence bands) is more complex.
For Bi$_2$Se$_3$ to be \textit{p}-type it has to be doped, \textit{e.g} with Ca, and Ca doping reduces carrier mobility.
Our Bi$_2$Te$_3$ crystals display exceptionally high mobilities, much higher than in Bi$_2$Se$_3$:Ca which will require further improvements in the initial materials' growth.  We also note
that for many applications of the remarkable properties of the topological Dirac states it is not necessary to stay at DP.  Indeed, to exploit the helical spin texture and strong spin polarization of the topological states one has to be above or below DP, since spins at the singular DP point are free to rotate \cite{SpinCusp-Lukas2014}.

\vspace{3mm}
\noindent \textbf{Supplementary Note 2\\
\noindent Evaluating energy barriers for defect migration.}\\
To evaluate the stability of compensating defects and to estimate energy barriers for defect migration, we performed isochronal anneals of a Bi$_2$Te$_3$ crystal irradiated with 2.5 MeV electrons at 20 K to a relatively low dose $\phi = 76~ \textrm{mC/cm}^2$ (below type conversion) and traced how the hole concentration $p$ evolved relative to the initial value, positing that the incremental change $\Delta p$ was proportional to the concentration of remaining defects. The initial decrease in hole concentration (here measured at 1 T) was from $3.36 \times 10^{18} \textrm{cm}^{-3}$ to $7.5\times 10^{17} \textrm{cm}^{-3}$ obtained after cycling to room temperature (RT) and back to 4.2 K.
Annealing was performed in 30 min steps each with the temperature increment of $5^\circ\textrm{C}$. After each annealing step, the sample was transferred into the cryostat for the measurements of resistivity and Hall effect at 4.2 K. Above $\sim 70^\circ\textrm{C}$ a measurable change in carrier density (shown in Supplementary Fig.~3a) becomes apparent, and above $\sim 150^\circ \textrm{C}$ the recovery is complete. For the determination of migration energy we used a standard method \cite{Damask-Dienes1963}: we calculated the increment of the carrier concentration change $\Delta p_i = p_i - p_{i-1}$ at each annealing step and normalized it to the remaining concentration of holes after the preceding step $p_{i-1}$, namely $\Delta p_i/p_{i-1}$. We assumed that the annealing process in each step obeys the rate equation $\textrm{d}p/\textrm{d}t = -K p$ corresponding to exponential decay law,
$p = p_\textrm{o} e^{-Kt}$, where $K$ is the diffusion coefficient $K \propto e^{-E_\textrm{b}/kT}$ with an energy barrier $E_\textrm{b}$ impeding the migration of defects.
The rate of annealing in step $i$, $\Delta p_i/p_{i-1}$ ($\approx \textrm{d}\textrm{ln} p/\textrm{d}t$) is proportional to $K$, thus the slope of $\textrm{ln}(\Delta p_i/p_{i-1}) \propto  \textrm{ln}(e^{-E_\textrm{b}/kT}$) \textit{vs}. $1/k_\textrm{B} T$ gives an estimate of the energy barrier controlling defect migration, $E_\textrm{b} \simeq 0.794~ \textrm{eV}$ (Supplementary Fig.~4b).  This is a typical value for the migration energy of vacancies and over an order of magnitude above $\lesssim 0.08~ \textrm{eV}$ expected for interstitials  \cite{Damask-Dienes1963}. The slow annealing process observed in a crystal irradiated to just above neutrality point in Fig.~3 of the main text is consistent with this large value of $E_\textrm{b}$; this explains why it takes hundreds of hours at RT to get back to the charge neutrality point. We note that at room temperature interstitials do not contribute since their concentration is about six orders of magnitude below that of vacancies \cite{Damask-Dienes1963}.

\vspace{3mm}
\noindent \textbf{Supplementary Note 3\\
\noindent Stability of surface states at CNP.}\\
The 2D transport at CNP will have contributions from the metallic Dirac bands and from the subsurface two-dimensional electron gas (2DEG) states arising from bending of bulk bands. We note that 2DEG states become more visible in transport when the Fermi level is in the bulk bandgap.  Charge transfer from the environment responsible for the band-bending would primarily be into the bulk bands nearest to the surface, if they are available. In the case of Bi$_2$Te$_3$, where Dirac point is in the valley of the bulk valence band (BVB) \cite{Zhang-NatPhys09} and where CNP crosses both Dirac and BVB, the charge transfer would be mostly into BVB and the band-bending is not a controlling factor. We should add that the subsurface 2DEG could be potentially controlled by the related ion implantation techniques, such as subsurface $\delta$-doping or 2D single-ion doping, well developed and fine-tuned in technology of semiconductor heterostructures.

We remark that the achieved high bulk resistivity values at CNP can be particularly useful in spintronic TI-based devices. A simple estimate based on spin-torque transfer induced by spin-polarized currents through topological surfaces indicates that in a typical thin layer structure spin-torque transfer (STT) will be hugely enhanced; indeed in our irradiated TIs even at room temperature it will be only $\sim 20-30\%$ below the maximum STT expected from the spin-polarized surfaces alone. This is a large improvement over the unirradiated TIs \cite{Ralph2014} where STT is ~70\% below maximum owing to the significant shunting by the bulk.

\vspace{3mm}
\noindent \textbf{Supplementary Note 4\\
\noindent Variable range hopping bulk charge transport at low carrier density.}\\
Sheet resistance $R_{\textrm{xx}}$ \textit{vs}. temperature of Bi$_2$Te$_3$ crystal at the charge neutrality point (shown in main Fig.~4a) at low temperatures is temperature independent, consistent with minimum conductance $G\cong 20~ G_0$ ($G_0 = e^2/h$ is the conductance quantum) of surface conduction channels, see main text. At higher temperatures two types of activated bulk behavior are observed. With a note of caution, since the temperature range is small here, at the highest temperatures (near room temperature) a simple activation law $R_{\textrm{xx}} \varpropto e^{{E_\textrm{a}}/{k_\textrm{B} T}}$
with a very small activation barrier $E_a \sim 7~\textrm{meV}$ fits best (see inset in Supplementary Fig.~11a). We note that the barrier energy $E_\textrm{a}$ appears much smaller than the bulk gap $E_\textrm{g} \simeq 200-300~\textrm{meV}$, and well below room temperature equivalent to $\sim 30~ \textrm{meV}$.

This simple activated behavior changes below $\sim 220~\textrm{K}$ into variable range hopping (VRH) of the Efros-Shklovskii type:
$R_{\textrm{xx}} = R_0 \textrm{exp} [(T_{0}/T)^{1/2}]$.
A fit of $R_{\textrm{xx}}$ to the VRH law ( Supplementary Fig.~11b)  gives $E_{\textrm{VRH}} = k_\textrm{B}T_{0} \sim 160~\textrm{meV}$ equivalent to $T_0 \sim 1880~\textrm{K}$.
Variable range hopping is a tunneling transport between separated charge puddles created by large charge fluctuations at CNP which are poorly screened, leading to unreliable estimates of the true carrier densities from Hall transport.
Such behavior is expected in theory \cite{Shklovskii2012}, where VRH temperature scale is given by $T_0 \propto 1/ \xi$, and $\xi$ is the localization length of states in the vicinity of the Fermi level measured
in the units of cube root of carrier density $n^{1/3}$.
Using our value of $T_0 $ from the fit to VRH law and carrier densities $n \sim 10^{18}~\textrm{cm}^{-3}$,
we obtain a rough estimate \cite{Shklovskii2012} of $\xi \sim 0.4 n^{-1/3} \approx 4~\textrm{nm}$. Owing to this remarkably long localization length VRH can dominate transport over a large temperature range, from  $T\simeq 220~ \textrm{K}$ down to the low-temperature plateau in $R_{\textrm{xx}}$.